\documentclass[lettersize,journal]{IEEEtran}
\usepackage{amsmath,amsfonts}
\usepackage{algorithmic}
\usepackage{algorithm}
\usepackage{array}
\usepackage[caption=false,font=footnotesize,labelfont=rm,textfont=rm]{subfig}
\usepackage{textcomp}
\usepackage{float}
\usepackage{url}
\usepackage{verbatim}
\usepackage{graphicx}
\usepackage{cite}
\usepackage{booktabs}
\usepackage{stfloats}
\usepackage{multirow}
\usepackage{threeparttable}
\usepackage{makecell}

\begin{document}

\title{RIMformer: An End-to-End Transformer for FMCW Radar Interference Mitigation}

\author{Ziang Zhang, Guangzhi Chen,~\IEEEmembership{Member,~IEEE}, Youlong Weng, \\Shunchuan Yang,~\IEEEmembership{Senior Member,~IEEE}, Zhiyu Jia and Jingxuan Chen

\thanks{This work is supported in part by the National Natural Science Foundation of China under Grant 62101020, Grant 62141405 and Grant 62293495 and Special Scientific Research Project of Civil Aircraft under Grant MJZ5-2N22. \textit{(Corresponding author: Guangzhi Chen)}.}
\thanks{Ziang Zhang, Guangzhi Chen and Youlong Weng are with School of Electronic and Information Engineering, Beihang University, Beijing, China and Hefei Innovation Research Institute of Beihang University, Hefei, China (e-mail: ziangzhang@buaa.edu.cn; dazhihaha@buaa.edu.cn; youlong\_weng@buaa.edu.cn).}
\thanks{Shunchuan Yang, Zhiyu Jia and Jingxuan Chen are with School of Electronic and Information Engineering, Beihang University, Beijing, China (email: scyang@buaa.edu.cn; feihong\_J@buaa.edu.cn; ChenJX@buaa.edu.cn).}}

\maketitle

\begin{abstract}
Frequency-modulated continuous-wave (FMCW) radar plays a pivotal role in the field of remote sensing. The increasing degree of FMCW radar deployment has increased the mutual interference, which weakens the detection capabilities of radars and threatens reliability and safety of systems. In this paper, a novel FMCW radar interference mitigation (RIM) method, termed as RIMformer, is proposed by using an end-to-end Transformer-based structure. In the RIMformer, a dual multi-head self-attention mechanism is proposed to capture the correlations among the distinct distance elements of intermediate frequency (IF) signals. Additionally, an improved convolutional block is integrated to harness the power of convolution for extracting local features. The architecture is designed to process time-domain IF signals in an end-to-end manner, thereby avoiding the need for additional manual data processing steps. The improved decoder structure ensures the parallelization of the network to increase its computational efficiency. Simulation and measurement experiments are carried out to validate the accuracy and effectiveness of the proposed method. The results show that the proposed RIMformer can effectively mitigate interference and restore the target signals.
\end{abstract}


\section{Introduction}
\begin{figure}[tb]
\centerline{\includegraphics[width=85mm]{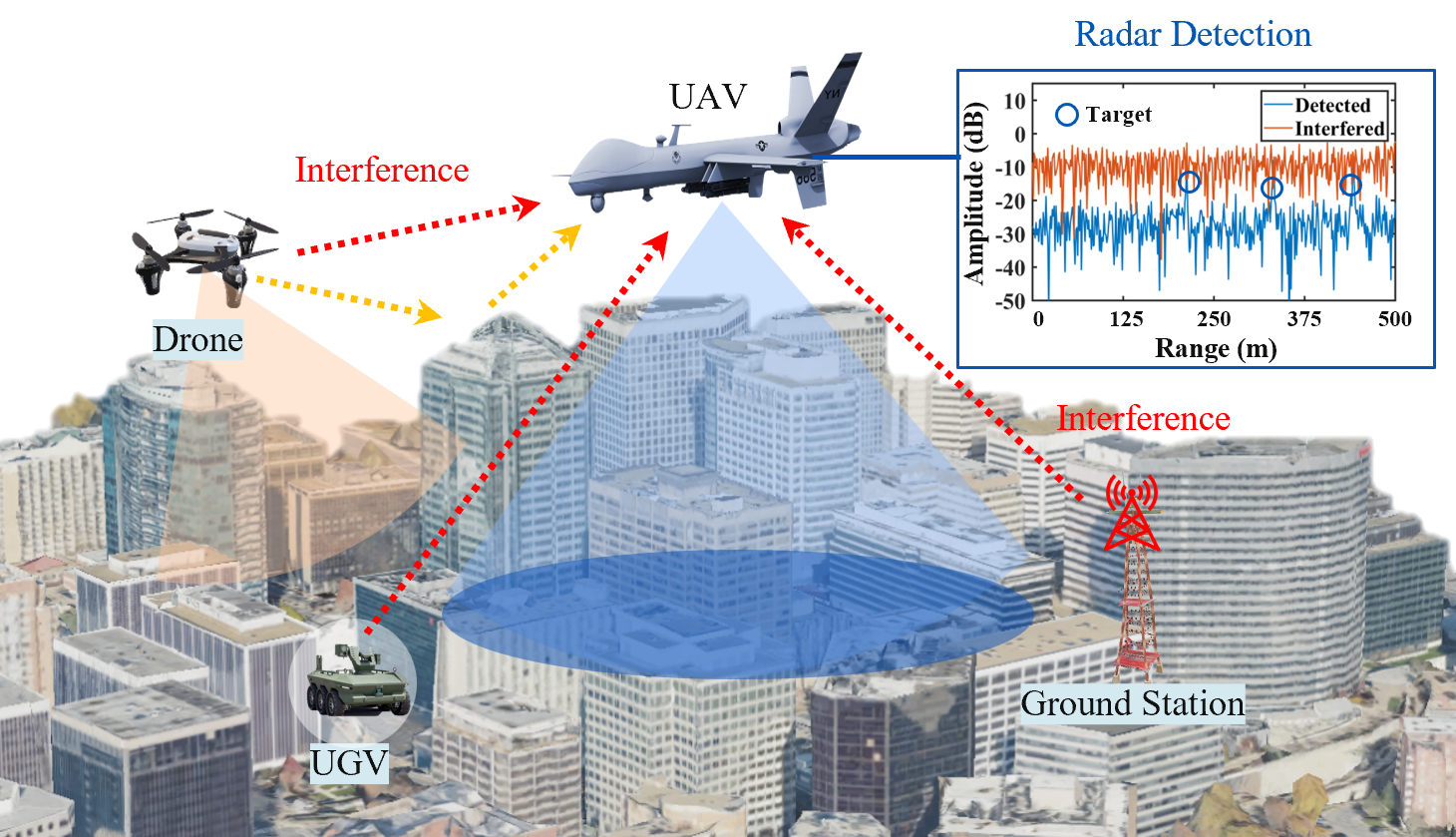}}
\caption{
Demonstration of a scenario where FMCW radar system on an UAV faces potential interference from other radars.
}
\label{street}
\end{figure}

\IEEEPARstart{W}{ith} the growth of the low altitude economy and smart city development, frequency-modulated continuous-wave (FMCW) radars are becoming increasingly important. FMCW millimeter-wave (mmWave) radar enhances the reliability of fully autonomous systems, including both unmanned aerial vehicles (UAVs) \cite{bg0} and unmanned ground vehicles (UGVs) \cite{bg1}. Compared with optical sensors, mmWave radars can provide accurate distance and velocity information, even under disturbances such as fog, heavy rain, and low lighting conditions \cite{bg2}. This capability is especially relevant for varieties of remote sensing applications, including height measurement \cite{drone}, radar imaging \cite{use1}, environment monitoring\cite{use2}, \cite{use3} and target detection \cite{use4}, \cite{use5}. With the growing adoption of FMCW radars in various transportation modes and ground monitoring stations, the electromagnetic interference among these systems has become a critical concern \cite{bg8}. As illustrated in Fig.~\ref{street}, the FMCW radar system discerns targets and obstacles by capturing reflected signals. However, the signals transmitted from other FMCW radar systems cause interference, which can surpass the power level of its own echo signal. Typically, interference in this context tends to raise the noise floor and diminish the visibility of targets. Furthermore, it obscures weak signals, consequently affecting detection capabilities and temporarily creating blind spots. Methods for effectively reducing interference have attracted considerable attention.

Various studies have been conducted on FMCW radar interference mitigation. Several methods have been proposed to either avoid or suppress interference through system designs. With a new orthogonal noise waveform design method \cite{sys1}, the noise waveform design of a radar system can be transformed into a phase recovery problem. In addition, strategies based on group delay filters \cite{sys2} and resource allocation methods \cite{sys3} based on time-frequency-division multiple access (TFDMA) can be used to mitigate vehicular radar interference. Another approach is to establish coordinated programs for different radars. RadarMAC \cite{xt1} includes a system architecture and a dynamic radar parameter assignment algorithm for mitigating radar interference in self-driving cars.

Other methods for mitigating interference include signal processing methods, which use a series of algorithms to identify and remove interference from radar signals. In \cite{sp1}, a wavelet denoising method was proposed to suppress interference from other vehicle signals in radar systems. The interference signal was extracted from the output of a time-domain low-pass filter via the wavelet transform and thresholding. An adaptive noise canceller (ANC) was designed for radar interference suppression \cite{sp2}. The interference produced by the attacking radar was cancelled from the main channel using the correlation between the positive and negative half spectra in the frequency domain. In \cite{sp3}, the interfered parts of the beat frequencies within a sweep were suppressed in the short-time Fourier transform (STFT) domain by beat frequency interpolation. An autoregressive (AR) model \cite{sp4} was used to reconstruct a disturbed baseband signal in the time domain. In addition, empirical mode decomposition (EMD) was used to conduct interference mitigation for the interfered signal \cite{sp5}. Signal separation \cite{sp6} was achieved by exploiting the sparsity differences between target signals and interference signals in the tunable Q-factor wavelet transform (TQWT) domain. After converting interference suppression into an optimization problem, robust adaptive beamforming \cite{sp7} and Hankel matrix decomposition \cite{sp8} were used to solve the problem. In \cite{sp9}, the RIM was achieved by extracting the target echoes with a row-sparse constraint.

In recent years, deep learning methods have been increasingly applied to FMCW radar interference mitigation tasks and have achieved impressive results. According to the input data, these methods can be divided into time-domain methods and time-frequency-domain methods. Among the former approaches, a gated recurrent unit (GRU) with self-attention \cite{dl1} was used to reconstruct time-domain signals. In some studies, time-domain signals were subjected to Fourier transforms and converted into two-dimensional data for processing with a convolutional neural network (CNN). A two-stage deep neural network (DNN) model with mask-gated convolution \cite{dl3} was proposed for radar interference detection and mitigation. In \cite{dl5}, a prior-guided method based on a complex-valued CNN was introduced to effectively eliminate interference in the time-frequency domain. In \cite{dilconv}, dilated convolution was used to achieve improved interference mitigation performance. Quantization techniques \cite{dl6} have been investigated to perform CNN-based denoising and interference mitigation on radar signals, resulting in reduced memory requirements. In \cite{unsuper}, a feature-oriented unsupervised adaptive suppression network was proposed to adaptively suppress the mutual interference between FMCW radars.

Recently, The Transformer \cite{dl9} has demonstrated impressive potential in the field of signal processing. This kind of network is particularly adept at capturing repetitive patterns with extended dependencies and is well suited for reconstructing periodic targets from disturbed signals. In this paper, a novel Transformer-based FMCW radar interference mitigation (RIM) method, named as the RIMformer, is proposed to mitigate interference and recover the target signal. The main contributions of this paper are summarized as follows.

1) A Transformer-based architecture is introduced to FMCW radar signal processing, which provides an effective RIM solution. To obviate the need for supplementary radar signal processing steps, the proposed RIMformer is designed with an end-to-end architecture to directly receive and generate signals in the time domain.

2) The encoding and decoding components in the RIMformer architecture are improved. Dual multi-head self-attention mechanisms using relative position coding and convolutional enhancement techniques are proposed to improve the performance of the network in terms of capturing features of beneficial components of interfered signals at different scales.

3) Both simulation and measurement experiments are carried out to validate the accuracy and effectiveness of the proposed RIMformer, which exhibits good performance in terms of its interference cancellation and target signal reconstruction capabilities.

This paper is organized as follows. In Section II, the proposed RIMformer for interference mitigation is introduced in detail. Section III contains a series of quantitative comparisons and analyses implemented through simulated experiments. Practical measurements are also carried out to validate the performance of the proposed method in Section IV. Finally, conclusions are drawn in Section V.

\begin{figure}[tb]
\centerline{\includegraphics[scale=0.68]{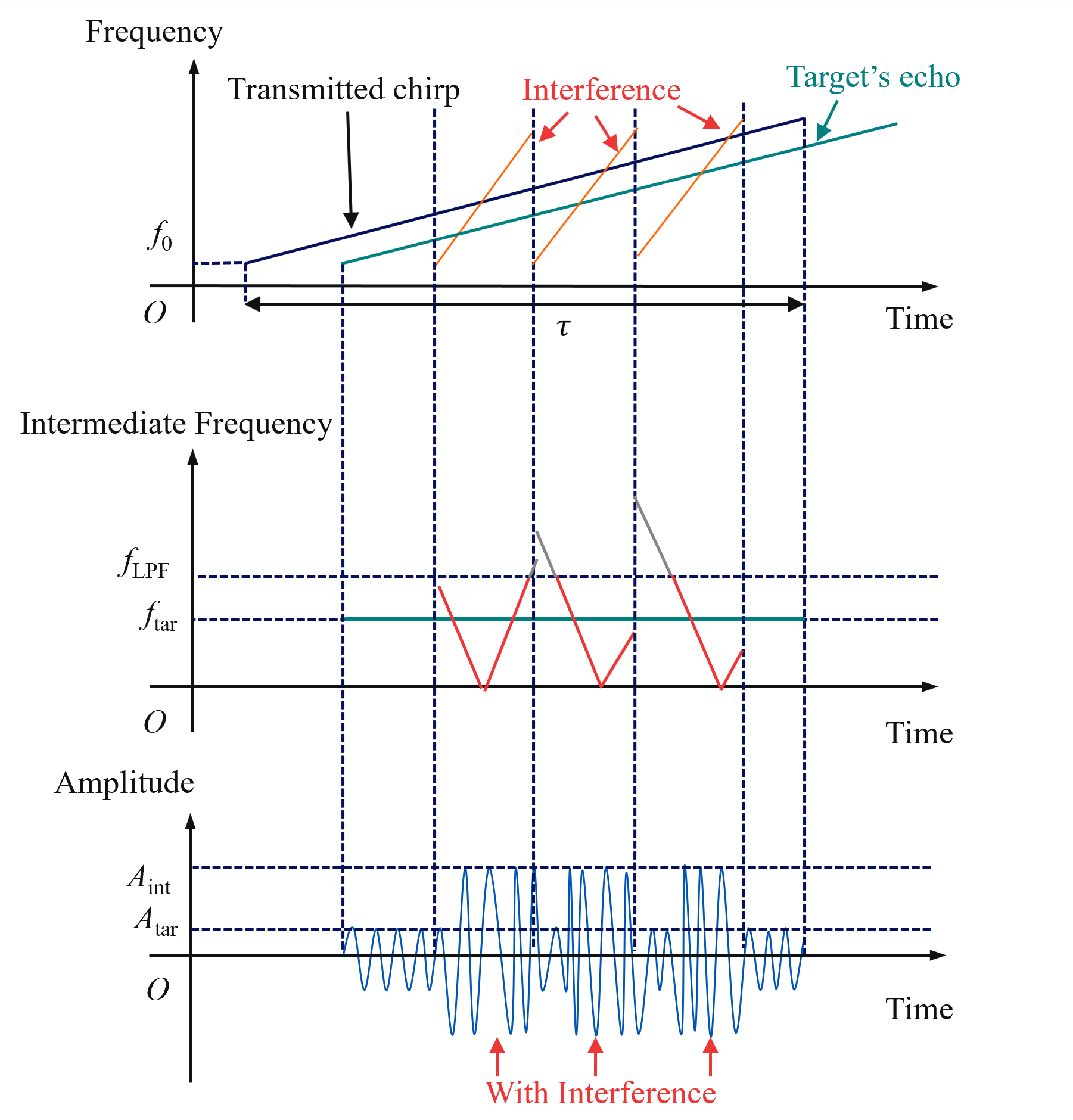}}
\caption{Chirp signals from other FMCW radars that are contained in the same frequency band are received, which cause the pulse-like interference in the time domain.}
\label{signal}
\end{figure}

\section{Methodology}

\subsection{FMCW Radar Interference}

\begin{figure*}[!t]
\centering
\includegraphics[width=180mm]{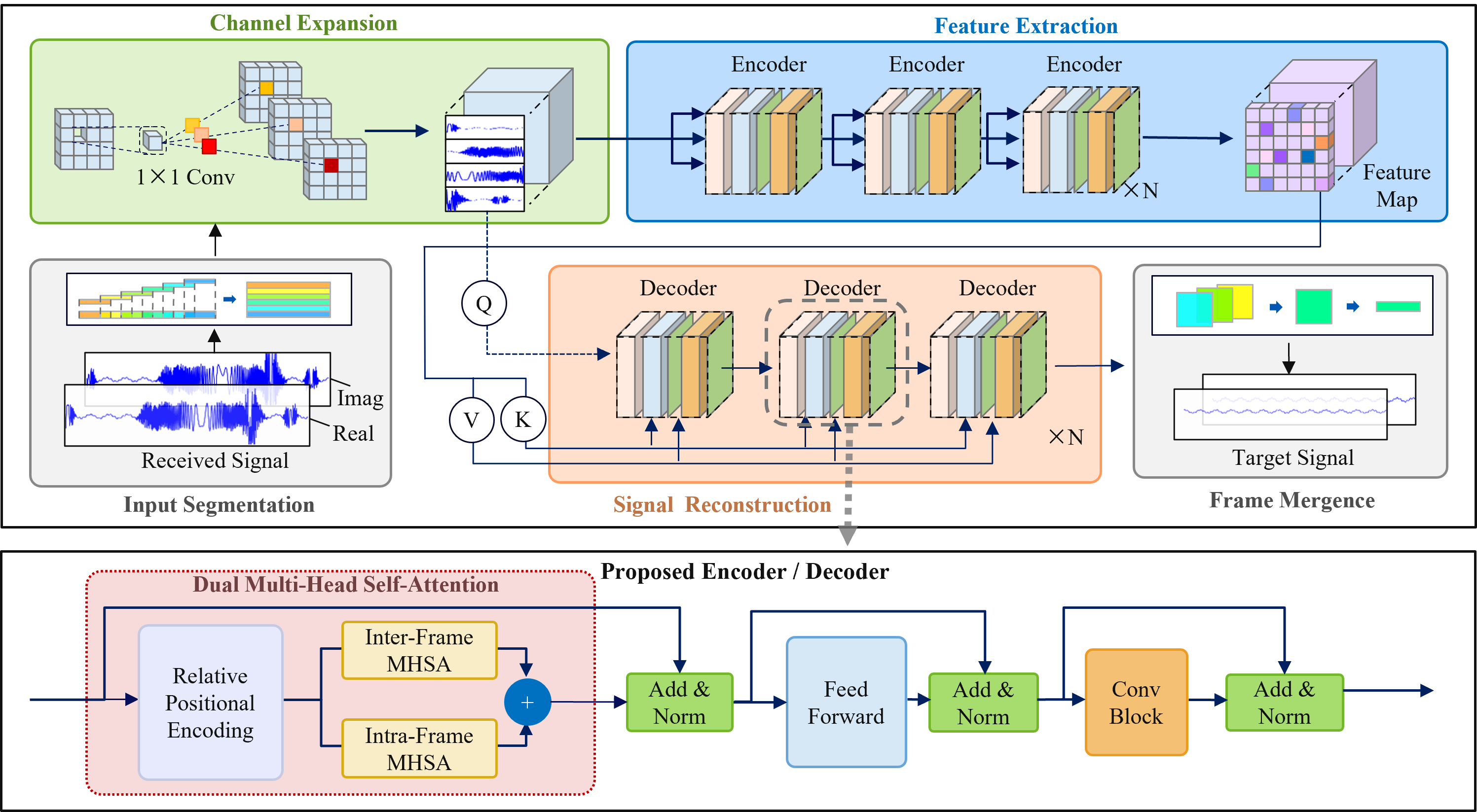}
\caption{The RIMformer adopts a typical encoder-decoder architecture. Within the decoders, the initial signal serves as the query, while the encoding outcome functions are the key and values. The results obtained from the decoders are amalgamated to form the ultimate output, which mirrors the shape of the input time-domain signal. Both the encoders and decoders exhibit identical structures, featuring key components such as dual multi-head self-attention mechanisms, feedforward mechanisms, and convolutional blocks.}
\label{overall}
\end{figure*}

In an FMCW radar remote sensing system, the received echo signal is mixed with the transmitted signal. Subsequently, low-pass filtering (LPF) is applied to generate an intermediate frequency (IF) signal. The signal passes through an analog-to-digital converter (ADC) and subsequent signal processing, which can obtain target parameters such as position, velocity and angle parameters. The transmitted signal in a chirp cycle can be expressed as
\begin{align}
    s_{\text{T}}\left(t\right)=A_{\text{T}}\exp{\left\{j 2\pi \left( f_0 t + \frac{K}{2} t^2 \right)\right\}},\ 0<t<T
\end{align}
where $f_0$ denotes the start frequency of the transmitted signal, $K=B/T_c$ denotes the frequency modulation (FM) slope of the transmitted signal, $B$ is the effective bandwidth, $T_c$ is the effective time width and $T$ is the time duration of the chirp signal. For each target, the echo signal can be expressed as:
\begin{equation}
s_r(t) = A_{\text{r}} \exp\left\{j 2\pi \left( f_0 (t - \tau) + \frac{K}{2} (t - \tau)^2 \right)\right\}
\end{equation}
where $\tau$ is the echo delay of the target. Notably, the signal encompasses not only the echo derived from the target under detection but also possibly includes interference signals from radar transmitters. Typically, the chirp slopes of interference signals from other radar systems differ from that of the transmitter. After conducting mixing and LPF, the time-domain expression for the signal with interference $s_{\text{int}}(t)$ can be derived as follows:
\begin{align}
\hat{s}(t)=\text{LPF}\left\{\left[s_{r}(t)+s_{\text {int }}(t)\right] \cdot s_{T}^{*}(t)\right\}
\end{align}
where the superscript * denotes the complex conjugate. The expression for the interfered IF signal ${\hat{s}}_{\text{IF}}(t)$ is obtained by summing (3) and the noise $s_{\text{n}}(t)$, which can be expressed as
\begin{align}
{\hat{s}}_{\text{IF}}\left(t\right)=\text{LPF}\left\{s_{r}(t) \cdot s_{T}^{*}(t)\right\}+\text{LPF}\left\{s_{\text{int}}(t) \cdot s_{T}^{*}(t)\right\}+s_{\text{n}}\left(t\right)
\end{align}

The interference formation process is illustrated in Fig.~\ref{signal}. After implementing mixing and LPF, the interference produces significant fluctuations in the time domain. $f_\text{LPF}$ is the bandwidth of the filter. $f_\text{tar}$ is the differential frequency corresponding to the target. $A_\text{tar}$ and $A_\text{int}$ are the amplitudes of the expected and interfered signals, respectively. The immediate consequence of interference is its broad spectrum, which extends across the entirety of the bandwidth. This spectrum broadening effect increases the level of background noise in the signal, diminishes the SINR of the target and even drowns the targets with the small radar cross section (RCS). Such interference significantly impairs the ability of the radar system to detect targets.

\subsection{The Proposed RIMformer}

\begin{figure*}[!t]
\centerline{\includegraphics[width=170 mm]{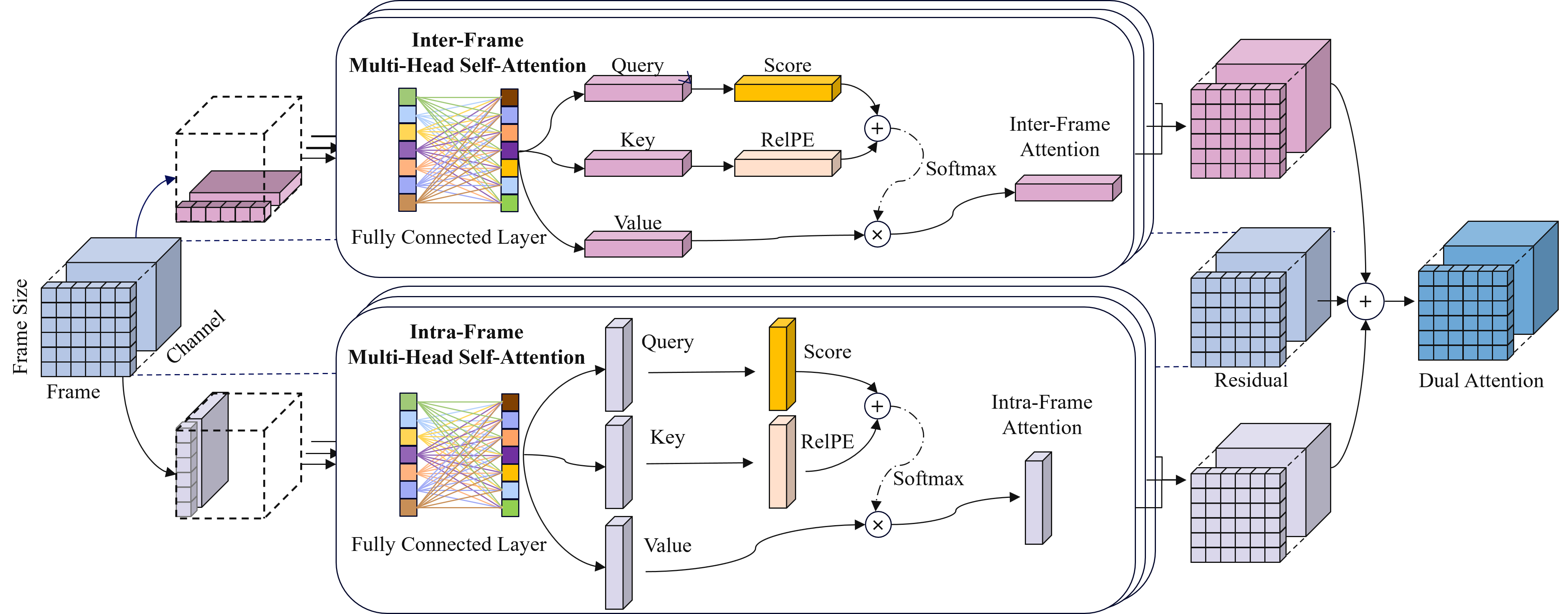}}
\caption{In the dual multi-head self-attention structure, the input data are split into different dimensions according to their inter-frame and intra-frame dimensions. Self-attention is then computed separately. The output is obtained by merging the two attention results and the residual.}
\label{dual}
\end{figure*}

Due to the exceptional sequence modeling capabilities, Transformer-based methods have been applied across a spectrum of signal processing tasks. The Transformer processes sequences with a self-attention mechanism that allows every element to directly participate in the handling of every other element. This mechanism enables the model to effectively capture long-range dependencies. We are inspired to harness the ability of the Transformer to capture long-range characteristics within time-domain signal sequences and reconstruct the targets from an interfered signal.

We propose the RIMformer based on the encoder-decoder Transformer structure, which is designed specifically for mitigating FMCW radar interference. The overall network framework is illustrated in Fig.~\ref{overall}. The network inputs time-domain signals within a chirp duration with interference from an FMCW radar system and reconstructs the target waveforms. An advantageous aspect of the end-to-end design is its capacity to guide the constructed model from the initial input to the ultimate output. This ability minimizes the of the network reliance on manual preprocessing and postprocessing steps. This approach affords the model great adaptability to the given data and diminishes the challenges associated with manual data processing and parameter acquisition tasks.

In the RIMformer, encoders are employed to perform feature extraction on the input interfered signal, while decoders utilize the encoded information for IF signal reconstruction purposes. As shown in Fig.~\ref{overall}, both the encoder and decoder share a common structure. In contrast to the original Transformer, the initial self-attention block in the decoder is omitted. This decision is grounded in the realization that the interference mitigation task does not necessitate the prediction of future data but focuses solely on restoring the data for the current time step based on contextual information. Consequently, both the encoder and decoder accept the raw data as inputs, thus ensuring the parallelism of the network. The results of the sequence of encoders serve as the key and query inputs in the attention computation of the decoders.

At the beginning of the processing flow of the RIMformer, the input interfered radar IF signal is decomposed into subsegments by a sliding window algorithm. These subsegments are then merged into two-dimensional data. This process is performed to facilitate dual attention calculations, which encompass attention at two different distance scales. After performing channel expansion via a one-dimensional convolution, the interrelationships within the interfered signal are captured by a series of encoders in the feature extraction stage. Upon combining the information derived from the feature map, the interference is removed, and the target is then reconstructed by multiple decoders. In the dual self-attention calculation, relative positional encoding is employed instead of absolute positional encoding. Additionally, the network incorporates supplementary convolutional block layers. This integration step capitalizes on the proficiency of Transformers in terms of capturing long-range relationships and the effectiveness of convolution at extracting localized features. Ultimately, the output of the decoders is consolidated into reconstructed one-dimensional signals with the same shape as that of the input radar IF signal.

In the preprocessing stage, the time-domain interfered signal samples acquired in each sweep are partitioned into several subsegments by means of a sliding window algorithm. These subsegments are spliced in new dimensions to generate two-dimensional data. The relationship between subsegment $y$ and the input signal samples $x$ can be represented as
\begin{equation}
\mathrm{y}_{k}=x[k L: (k+1) L+M], k=0,1,2, \ldots, n
\end{equation}

\noindent where L is the distance of each slide of the window and $L+M$ is the length of a subsegment.

To enhance the ability of the network to capture information across varying distances, we propose a dual multi-head self-attention mechanism. In the context of self-attention, the allocation of weights to each input hinges upon the interactions between the inputs. An internal voting process is performed to ascertain which inputs among them merit attention. After preprocessing, the interfered IF signals of the FMCW radar are converted to a structured format (frame, size, channel). The dual attention computation occurs in two distinct phases. As shown in Fig.~\ref{dual}, the input signals are partitioned by their dimensions, and attention is computed within each frame as well as between frames. Subsequently, the outputs of these two attention mechanisms are amalgamated. The intra-frame attention delineates the pertinence of signals in the short term, while the inter-frame attention encapsulates long-term effects. Moreover, inspired by \cite{RPE}, relative position encoding is used, and the attention expression can be calculated as
\begin{equation}
\operatorname{Attention}(Q, K, V)=\operatorname{softmax}\left(\frac{Q K^{T}+S^{r e l}}{\sqrt{d_k}}\right) V
\end{equation}

\noindent where $S^{rel}$ is a variable representing the relative positions of elements, which can be learned during the training process. The input consists of queries ($Q$) and keys ($K$) with $d_k$ dimensions and values ($V$) with $d_v$ dimensions. They are calculated by learnable linear transformations, which can be presented as $Q_{i}=X W_{i}^{Q}$, $K_{i}=X W_{i}^{K}$, and $V_{i}=X W_{i}^{V}$, where $W^*$ denotes the parameter matrix. Multi-head attention allows the model to jointly attend to information derived from different representation subspaces at different positions.
\begin{equation}
\begin{aligned}
\operatorname{MultiHead}(Q, K, V) & \!=\!\operatorname{Concat}\left(\text{head}_{1},\ldots, \text{head}_{\mathrm{h}}\right)W^{O}
\end{aligned}
\end{equation}
where
\begin{equation}
\begin{aligned}
\text{head$_i$} & \!=\!\text{Attention}_{\text{inter}}\left(Q_{i}, K_{i}, V_{i}\right)
\end{aligned}
\end{equation}

The inter-frame multi-head self-attention and intra-frame multi-head self-attention mechanisms are connected by residuals and finally combined via a summation operation.

\begin{figure}[tb]
\centerline{\includegraphics[scale=0.45]{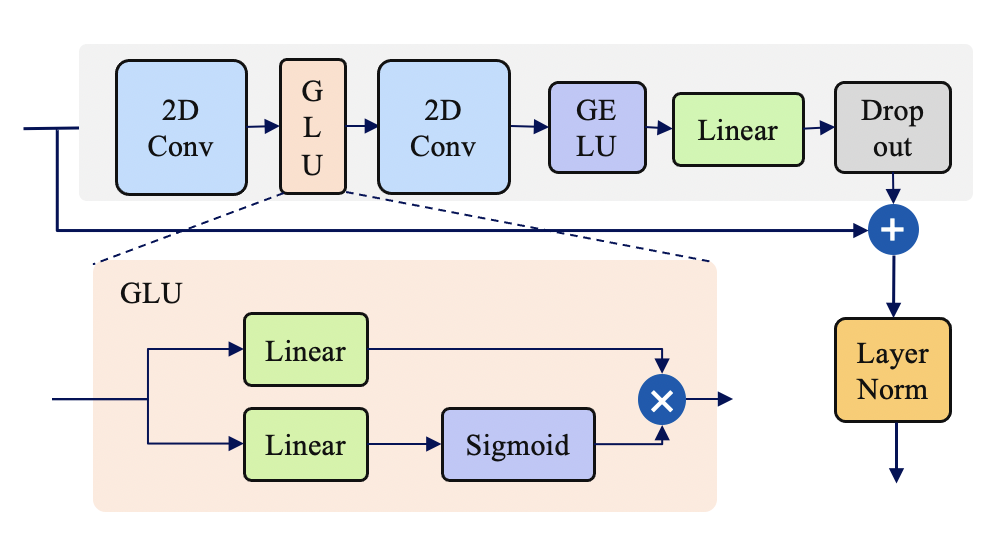}}
\caption{Convolutional block and details of the gated linear unit (GLU).}
\label{conv}
\end{figure}

In the RIMformer, convolution is employed to discern the interrelations among the elements situated in diverse locations within adjacent subsegments. To integrate the local feature utilization proficiency of a CNN with the ability of a Transformer to capture long-distance relationships, we incorporate a convolutional block into both the encoders and decoders. As shown in Fig.~\ref{conv}, inspired by \cite{conf}, the convolutional block encompasses convolutional layers, linear layers, and various activation functions. Positioned between these two convolutional layers is the gated linear unit (GLU) proposed by \cite{glu}, where two linear projections of \textbf{x} undergo pointwise multiplication, with one projection initially passing through the Sigmoid function.

\begin{equation}
    \text{GLU}(X)=(X * W+b) \otimes \sigma(X * V+c)
\end{equation}

At the end of the network, the outputs of the decoders are merged, and the resulting shape is consistent with that of the input signal. This step completes the end-to-end process of reconstructing the IF radar signal. The algorithm can be described as follows:
\begin{align}
\tilde{Y}[k L:(k+1) L]=
\begin{cases}
\frac{1}{2}\left(y_{k-1}[L:L+M]+y_{k}[0: M]\right) \\
\oplus y_{k}[M: L] , \quad k=1,2, \ldots, n \\
\vspace{-1ex}\\
y_{k}[0: L] , \quad k=0
\end{cases}
\end{align}
\noindent where $y_k$ is the decomposed subsegment in the output of the decoder and $\tilde{Y}$ is the merged result. The symbol $\oplus$ represents the operation of splicing two pieces of data.

Since the commonly used mean squared error (MSE) loss function can only reflect the differences among the predicted signals in the time domain, we propose a hybrid time-frequency loss function. This loss function is formulated to include the errors of the reconstructed signal in both the temporal and frequency domains. The time-domain loss and the frequency-domain loss capture the signal differences over different domains. This allows the network to not only focus on the signal values at a specific point in time but also to optimize the frequency-domain characteristics of the signal. Compared with the MSE loss, the proposed loss function can better represent the physical meaning of the IF signal and is helpful for achieving network convergence. The hybrid time-frequency loss function can be calculated as
\begin{equation}
\mathcal{L}_{T\&F}=\frac{1-\lambda}{\sqrt{N}} \left \|{Y-\widetilde{Y} }\right \|_2 + \frac{\lambda}{\sqrt{N}} \left \|{\mathcal{F}(Y)-\mathcal{F}(\widetilde{Y})}\right \|_2
\end{equation}

\noindent where $N$ is the length of the signal sequence and $\mathcal{F}$ denotes the spectrum of the signal after undergoing a Fourier transformation. $\lambda$ is a hyperparameter that is used to adjust the weights of the two components and is set to 0.3 based on experience.

\section{Numerical Experiments}
This section begins by describing the basic configuration of the experiments and the specific details of the training process. Then, ablation experiments are carried out to understand the role of each component in the proposed RIMformer. An exploration of the metric shifts that occur throughout the training process is undertaken to discern the optimal hyperparameters for the developed network. Finally, comparative experiments are conducted to evaluate the performance of our method and various comparative methods on a test dataset.

\subsection{Experimental Setup}

\begin{table}[tb]
    \caption{Simulation Parameters}
    \centering
    \renewcommand\arraystretch{1.5}
    \begin{threeparttable}
    \resizebox{\linewidth}{!}{
    \begin{tabular}{ccccc}
    \Xhline{1.5pt}
    \textbf{Type} & \textbf{Parameter} & \textbf{Vaule} & \textbf{Parameter} & \textbf{Value}\\
    \Xhline{1.5pt}
    \multirow{3}*{Victim} & StartFreq (GHz) & 76.5  & Samples & 1024\\
    ~ & Duration (s)& 20e-6  & Chirps & 128\\
    ~ & Slope (GHz/us) & 0.03  &  & \\
    \Xhline{1pt}
    \multirow{2}*{Target} & Number & (1,3) & Speed (m/s)& (3,45)\\
    ~ & Range (m)& (3,45) & Amplitude* & (0.4,3)\\
    \Xhline{1pt}
    \multirow{2}*{Interference} & Number & (0,5) & Slope (GHz/us)& (-0.0675, 0.0675)\\
    ~ & StartFreq (GHz)& (76.2, 76.8) & Amplitude* & (6,33)\\
    \Xhline{1.5pt}
    
    \end{tabular}}
     \begin{tablenotes}
      \footnotesize
      \item[*] The amplitudes are relative values.
     \end{tablenotes}
    \end{threeparttable}
    \label{sim}
\end{table}

A simulation dataset is constructed with reference to the existing FMCW mmWave radars and possible radar detection scenarios. As Table \ref{sim} demonstrates, multiple interference signal and target factors are considered. The simulated dataset consists of 8,000 pairs of interference and clean samples. The training set, validation set and test set are randomly divided at a ratio of 8:1:1. The number of samples per signal is 1024. Before training the RIMformer model, data preprocessing is implemented to optimize the learning process of the network. To enhance the numerical stability of the network and expedite its convergence process during training, a normalization step is applied to the input data. The sliding window algorithm is employed to segment the sequences into overlapping windows. The algorithm generates overlapping segments with lengths of 32 through a sliding distance of 16.

In this paper, three performance metrics are employed. First, the MSE serves as an indicator of the accuracy of the network in terms of reconstructing the echo signal. A small MSE implies a heightened degree of time-domain similarity between the reconstructed and original signals. Another metric for reflecting the efficacy of radar interference mitigation is the SINR, which reflects the visibility of the target in the frequency domain. A large SINR denotes improved target recognition. Additionally, considering the real-time requirements of sensors, the processing time required for single-frame data is a pivotal reference indicator for assessing system performance. The SINR is defined as the ratio of the mean power across the target peaks and the mean power of the interference and noise.

\begin{equation}
\operatorname{SINR}=10 \log10 \left(\frac{\frac{1}{\#_{\mathcal{T}}} \sum_{i \in \mathcal{T}}|s[i]|^{2}}{\frac{1}{\#_{\mathcal{N}}} \sum_{i \in \mathcal{N}}|s[i]|^{2}}\right)
\end{equation}
where $i$ is the index of the signals $s$, $\mathcal{T}$ is the set of target peaks, $\mathcal{N}$ is the set of background noise and noise caused by interference, and $\#$ denotes the cardinality of a set.

The code is implemented using the Python language and the PyTorch framework. The training and inference processes are carried out on an Ubuntu server with an Nvidia RTX 3090 graphics processing unit (GPU). The parameters of the RIMformer model are initialized using the Kaiming initialization method \cite{kaiming}. The adaptive moment estimation (Adam) optimizer \cite{adam} is employed to update the model parameters during the training process. The initial learning rate is set to $1\times10^{\-4}$. To dynamically adjust the learning rate during the training process, cosine annealing and warm restart \cite{warm} functions from the PyTorch library are utilized.

\begin{table}[tb]
    \caption{Results of Ablation Studies}
    \centering
    \renewcommand\arraystretch{1.2}
    \begin{threeparttable}
    \resizebox{\linewidth}{!}{
    \begin{tabular}{ccccc}
    \Xhline{1.5pt}
    \textbf{Model} & \textbf{Epochs} & \textbf{Training Loss\tnote{*}} & \textbf{MSE (e-5)\tnote{*}} & \textbf{Avg SINR (dB)}\\
    \Xhline{1.5pt}
    \multirow{5}*{RIMformer} & 100 & 0.036 & 4.001 & 5.462\\
    ~ & 200 & 0.024 & 2.163 & 21.944\\
    ~ & 300 & 0.016 & 2.362 & 22.399\\
    ~ & 400 & 0.015 & 2.139 & 23.348\\
    ~ & 500 & \textbf{0.014} & \textbf{1.879} & \textbf{23.878}\\
    \Xhline{1pt}
    \multirow{5}*{\shortstack{RIMformer \\ \quad- Dual Attention}} & 100 & 0.048 & 5.895 & -2.250\\
    ~ & 200 & 0.035 & 3.841 & 6.131\\
    ~ & 300 & 0.029 & 3.673 & 6.394\\
    ~ & 400 & 0.023 & 3.039 & 14.770\\
    ~ & 500 & 0.022 & 2.888 & 15.924\\
    \Xhline{1pt}
    \multirow{5}*{\shortstack{RIMformer \\ \quad- Conv Block}} & 100 & 0.037 & 3.984 & 4.427\\
    ~ & 200 & 0.023 & 2.154 & 21.529\\
    ~ & 300 & 0.021 & 2.098 & 21.613\\
    ~ & 400 & 0.017 & 1.882 & 22.698\\
    ~ & 500 & 0.015 & 2.014 & 22.763\\
    \Xhline{1pt}
    \multirow{5}*{\shortstack{RIMformer \\ \quad- Dual Attention \\ \quad- Conv Block}} & 100 & 0.034 & 3.967 & 12.623\\
    ~ & 200 & 0.026 & 3.402 & 12.279\\
    ~ & 300 & 0.024 & 3.343 & 14.017\\
    ~ & 400 & 0.021 & 3.638 & 14.132\\
    ~ & 500 & 0.020 & 3.511 & 14.410\\
    \Xhline{1pt}
    \multirow{5}*{\shortstack{RIMformer \\ \quad- $\mathcal{L}_{T\&F}$}} & 100 & - & 2.916 & 18.226\\
    ~ & 200 & - & 2.394 & 19.843\\
    ~ & 300 & - & 1.955 & 22.071\\
    ~ & 400 & - & 2.202 & 22.057\\
    ~ & 500 & - & 2.717 & 20.072\\
    \Xhline{1.5pt}
    
    \end{tabular}}
     \begin{tablenotes}
      \footnotesize
      \item[*] To avoid the effect of utilizing the magnitudes of the values, their \\ normalized values are used for the calculations.
     \end{tablenotes}
    \end{threeparttable}
    \label{ablation}
\end{table}

\subsection{Ablation Studies} 

 The encoder-decoder structure of the RIMformer integrates dual multi-head self-attention and convolution blocks. To better understand the role of each component, ablation experiments are performed to systematically isolate and evaluate the impact of the dual multi-head self-attention mechanism and the convolution blocks. To assess the contribution of the dual multi-head self-attention mechanism to the RIMformer model, this component is replaced with a normal self-attention mechanism. For the convolutional blocks, ablation experiments are conducted by excluding the convolution blocks from the model. To verify the effect of the joint frequency loss, the conventional MSE is used as a loss function for comparison. The model is trained with the same training parameters and evaluated based on the training loss, MSE and SINR.

The results obtained from the experiments are summarized in Table \ref{ablation}. After 500 epochs of training, the RIMformer achieves a training loss of 0.014, an MSE of 1.879, and an average SINR of 23.878 dB. When the dual multi-head self-attention mechanism is removed, the performance metrics exhibit notable changes. When the training loss increases to 0.022, the MSE increases to 2.888, and the SINR decreases to 15.924. When the convolution blocks are omitted, the SINR decreases to 22.763 dB. After removing both the dual multi-head self-attention mechanism and the convolution blocks, the combined absence of these structures results in a further performance decrease, with the SINR reaching 14.410 dB.

The dual multi-head self-attention mechanism plays a crucial role in capturing long-range dependencies and contextual information. The convolution blocks contribute to the ability of the RIMformer to extract local features. A comparative analysis of the ablation experiments confirms that the dual multi-head self-attention mechanism and convolution blocks contribute distinctively to the RIMformer model. Removing each structure causes the observed performance to degrade. When the hybrid time-frequency loss function is replaced with the MSE, the SINR reaches 22.071 dB after 300 epochs of training and then falls to 20.072 dB. These observed decreases in the performance metrics suggest that the new loss function enables the network to learn the information contained in the signal frequencies by employing.

\begin{figure}[tb]
\centerline{\includegraphics[scale=0.58]{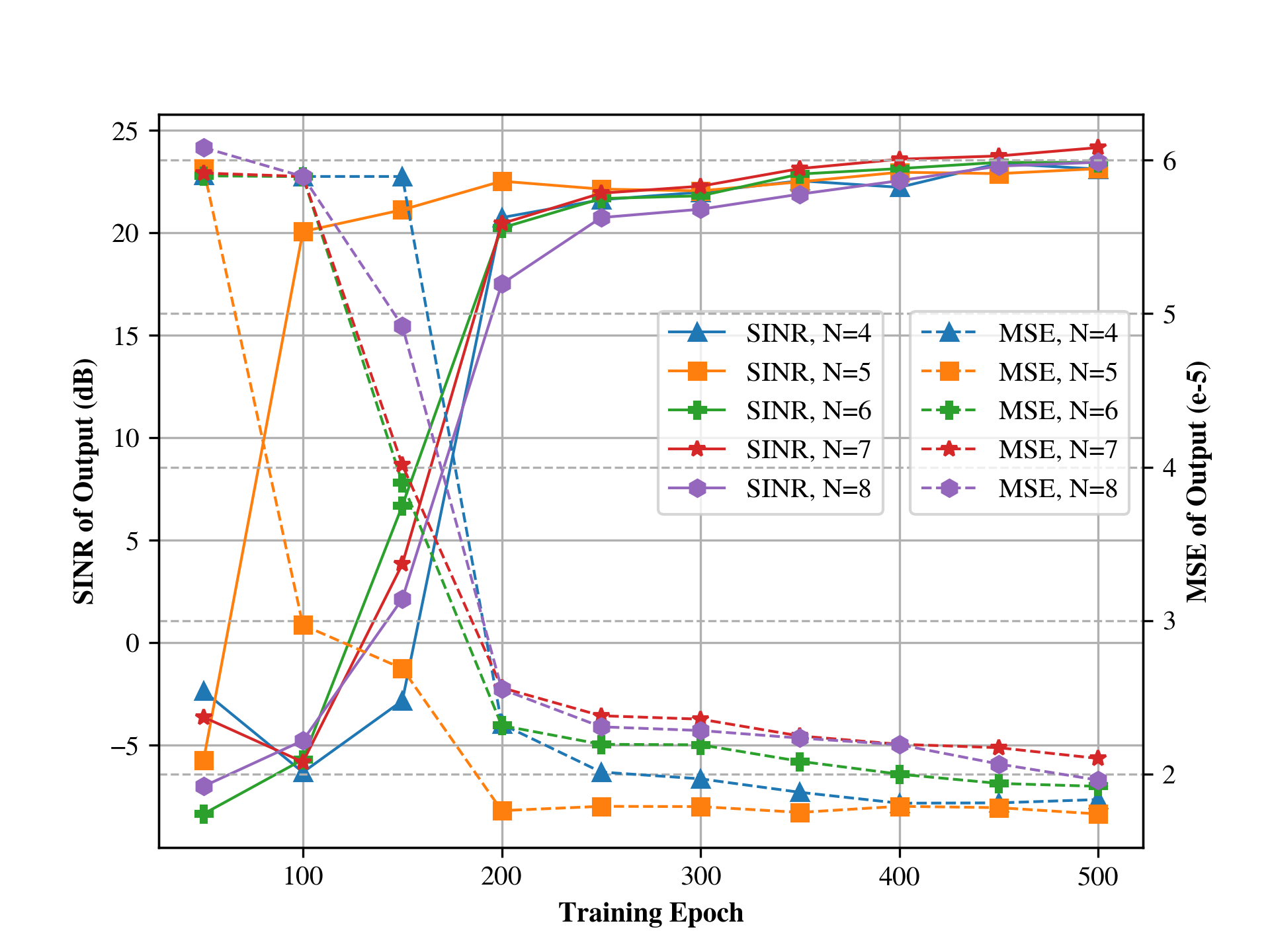}}
\caption{Changes exhibited by the MSE and SINR on the test set with increasing training epochs.}
\label{line-nlayer}
\end{figure}

The number $N$ of encoders and decoders in the network is a parameter that is subject to optimization. Comparative experiments are conducted to elucidate the relationship between the interference mitigation effectiveness of the network and the number of training epochs across various layer configurations. The outcomes, including the MSE and SINR produced for the target, are depicted in Fig.~\ref{line-nlayer}. The network demonstrates greater training ease with fewer encoders and decoders, converging within the first 200 epochs. As $N$ increases, the predictive accuracy of the network improves. This advancement comes at the cost of requiring more iterations for convergence, and it increases the susceptibility of the model to overfitting. Notably, at $N=8$, the network fails to converge even after 500 training rounds, which indicates an underfitted state. This phenomenon demonstrates the tradeoff between network depth and performance. Deeper networks tend to exhibit superior performance at the expense of heightened training complexity, an increased risk of overfitting, and elevated computational demands. The $N$ parameter of the RIMformer is set to 7 to ensure its interference mitigation performance and temporal efficiency. In addition, it can be seen from the figure that a low MSE in the time-domain signal does not necessarily imply a high SINR. therefore MSE cannot be used as the only evaluation metric for network training.

\begin{figure} [t]
	\centering
	\subfloat[\label{stft:a}]{
		\includegraphics[scale=0.20]{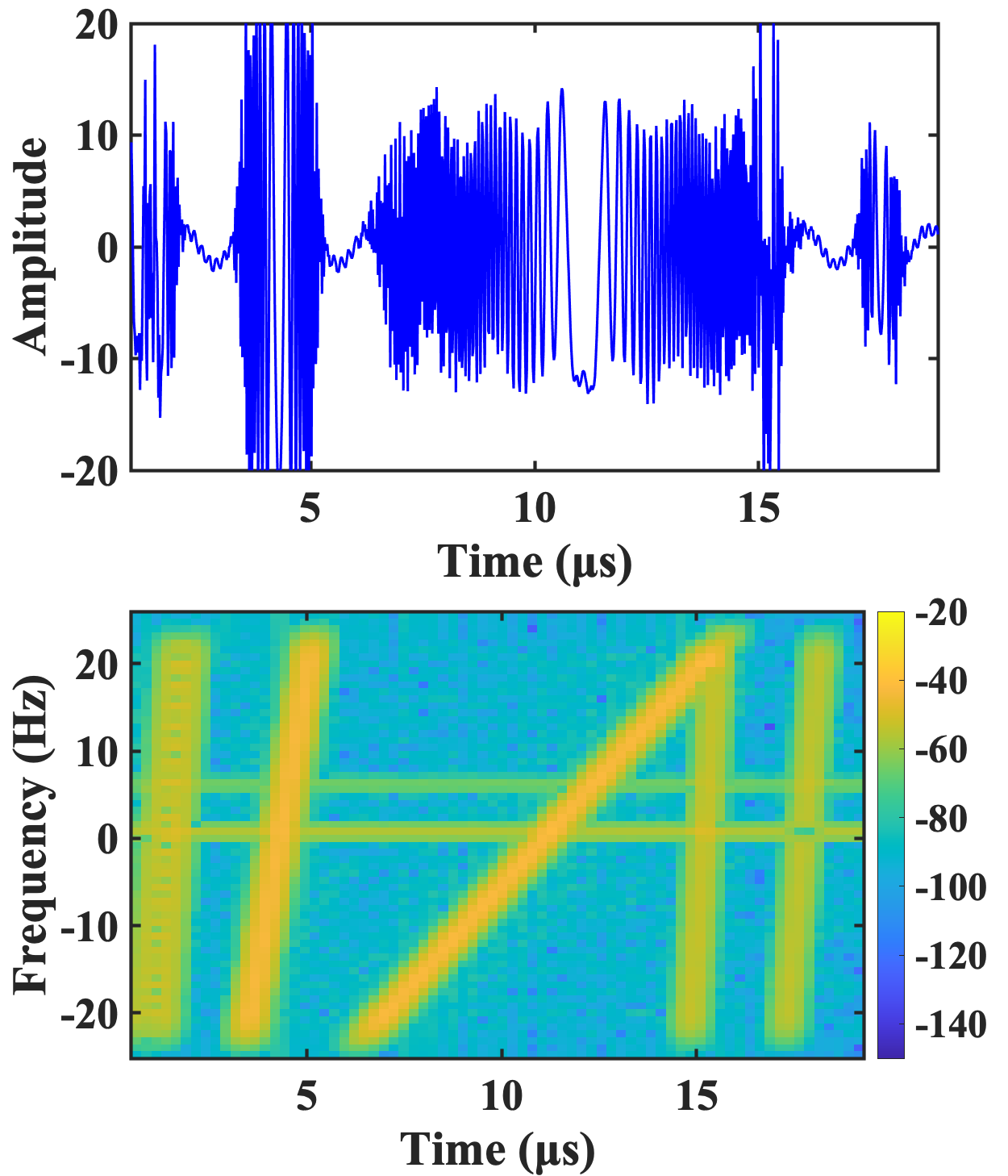}}
        \hspace{-2mm}
	\subfloat[\label{stft:b}]{
		\includegraphics[scale=0.20]{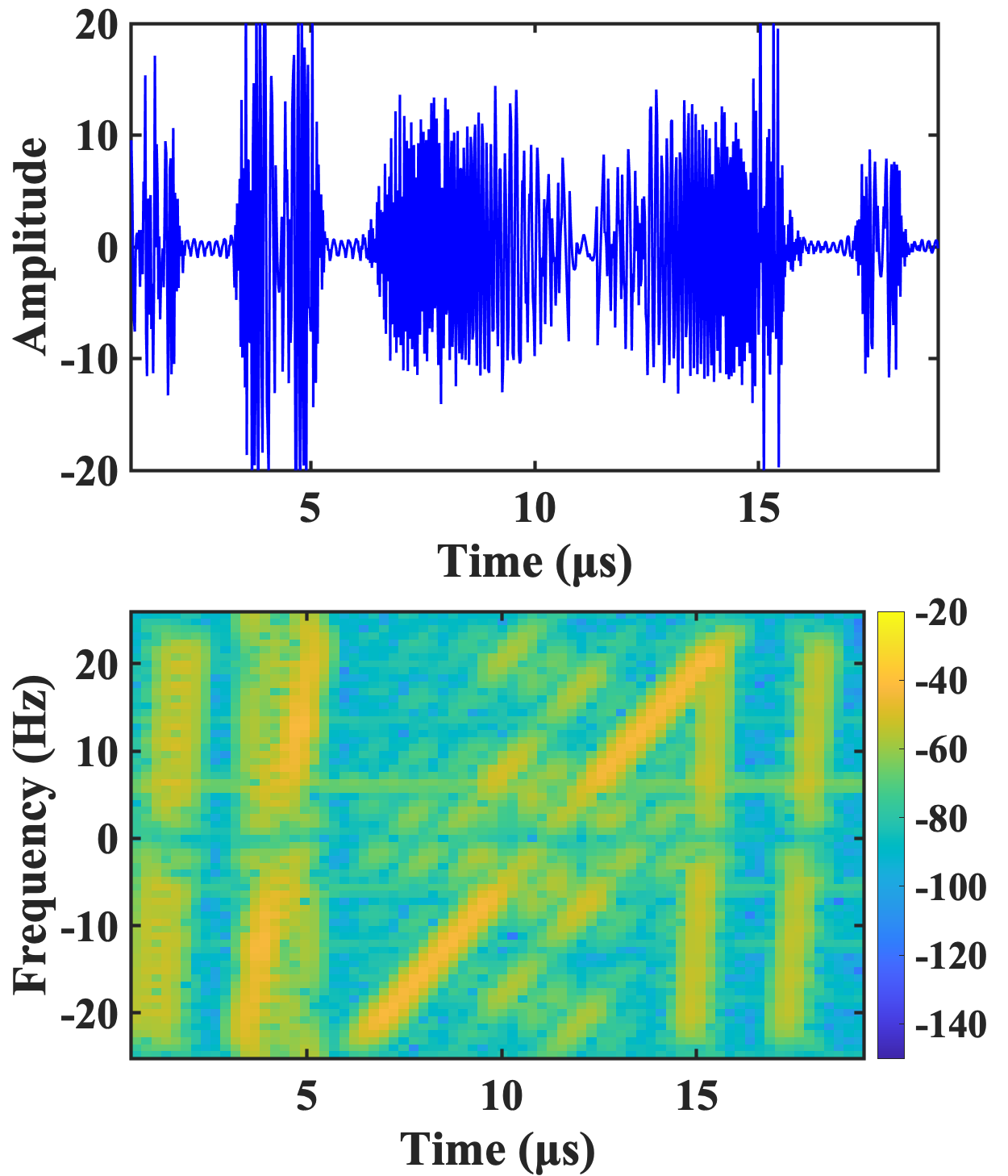}}
        \vspace{-3mm}
        \\
	\subfloat[\label{stft:c}]{
		\includegraphics[scale=0.20]{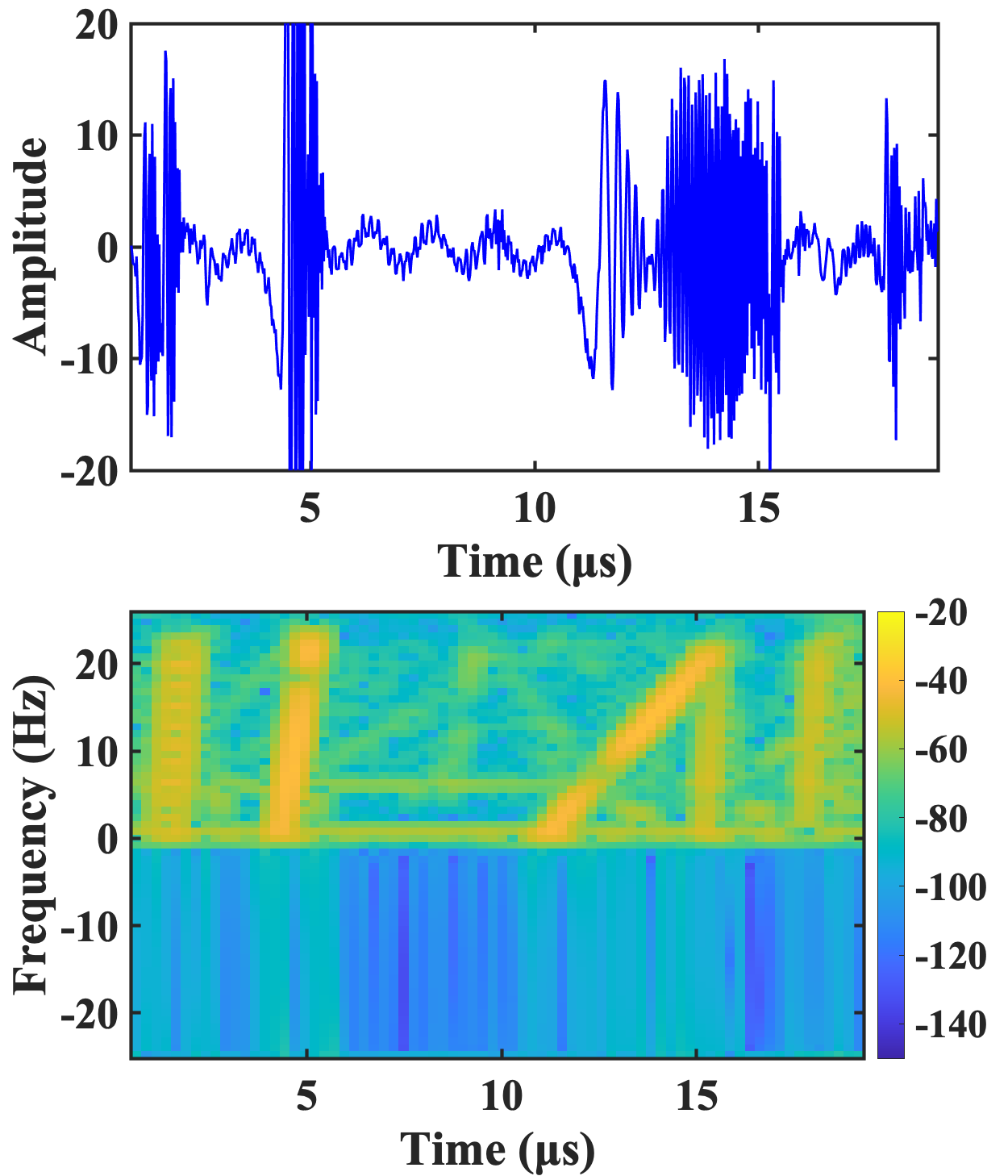}}
        \hspace{-2mm}
	\subfloat[\label{stft:d}]{
		\includegraphics[scale=0.20]{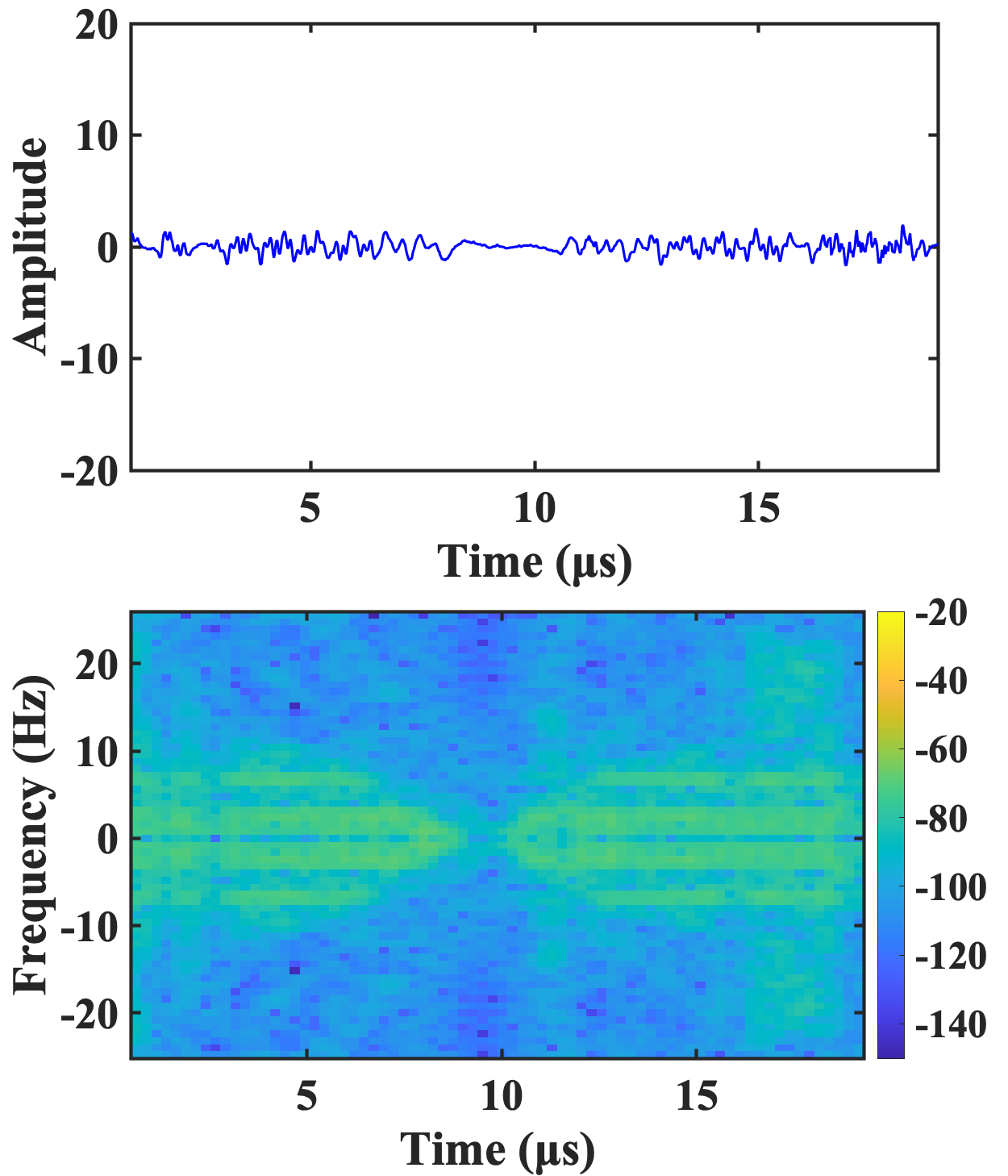} }
        \vspace{-3mm}
        \\
        \subfloat[\label{stft:e}]{
		\includegraphics[scale=0.20]{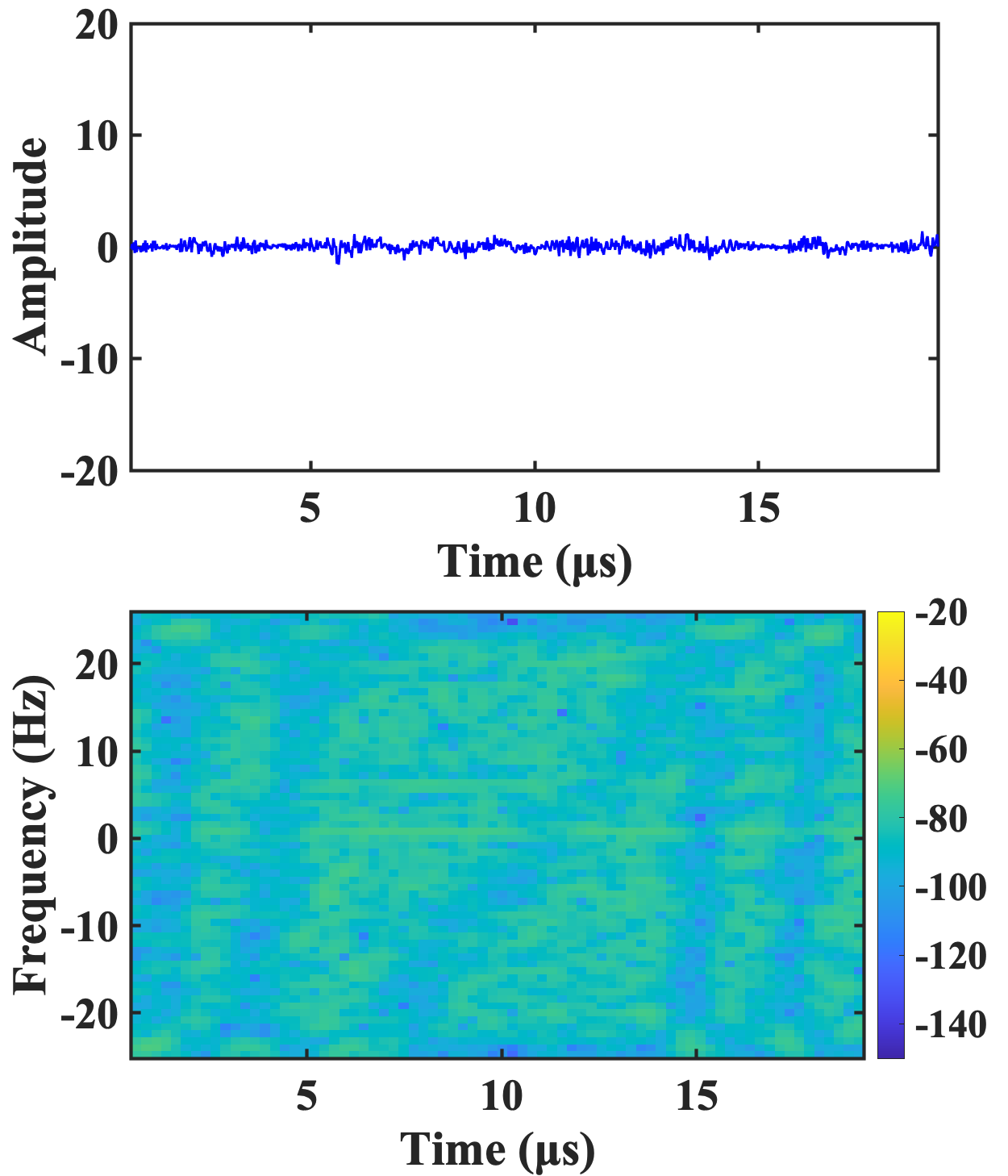} }
        \hspace{-3mm}
        \subfloat[\label{stft:f}]{
		\includegraphics[scale=0.20]{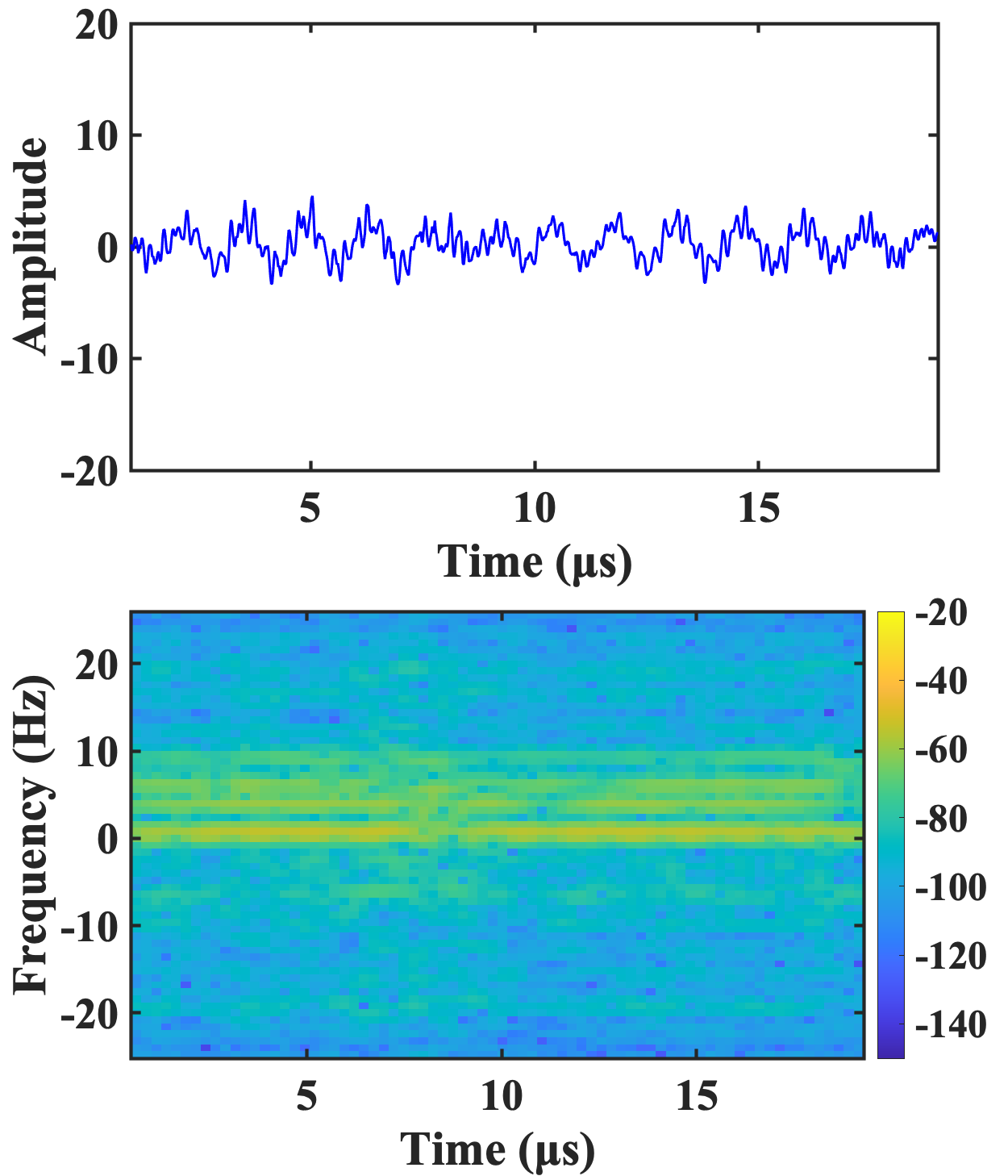} }
	\caption{Results produced by (a) received signal with interference, (b) wavelet denoising, (c) ANC, (d) attention-based BiGRU, (d) complex-valued CNN and (f) RIMformer for one of the simulated chirps, which are presented in the time and STFT domains. }
	\label{stft} 
\end{figure}

\begin{figure*} [t!]
	\centering
	\subfloat[\label{rd:a}]{
		\includegraphics[scale=0.28]{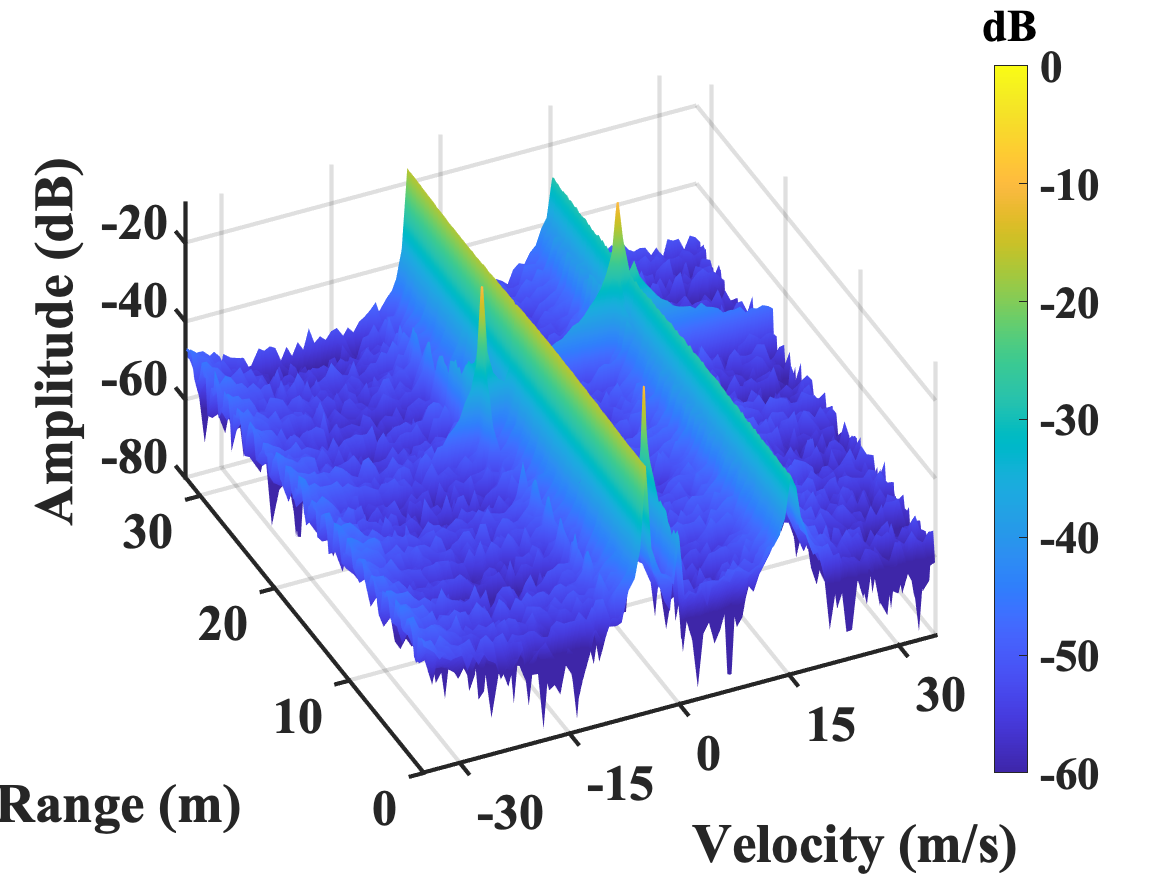}}
        \hspace{0mm}
	\subfloat[\label{rd:b}]{
		\includegraphics[scale=0.28]{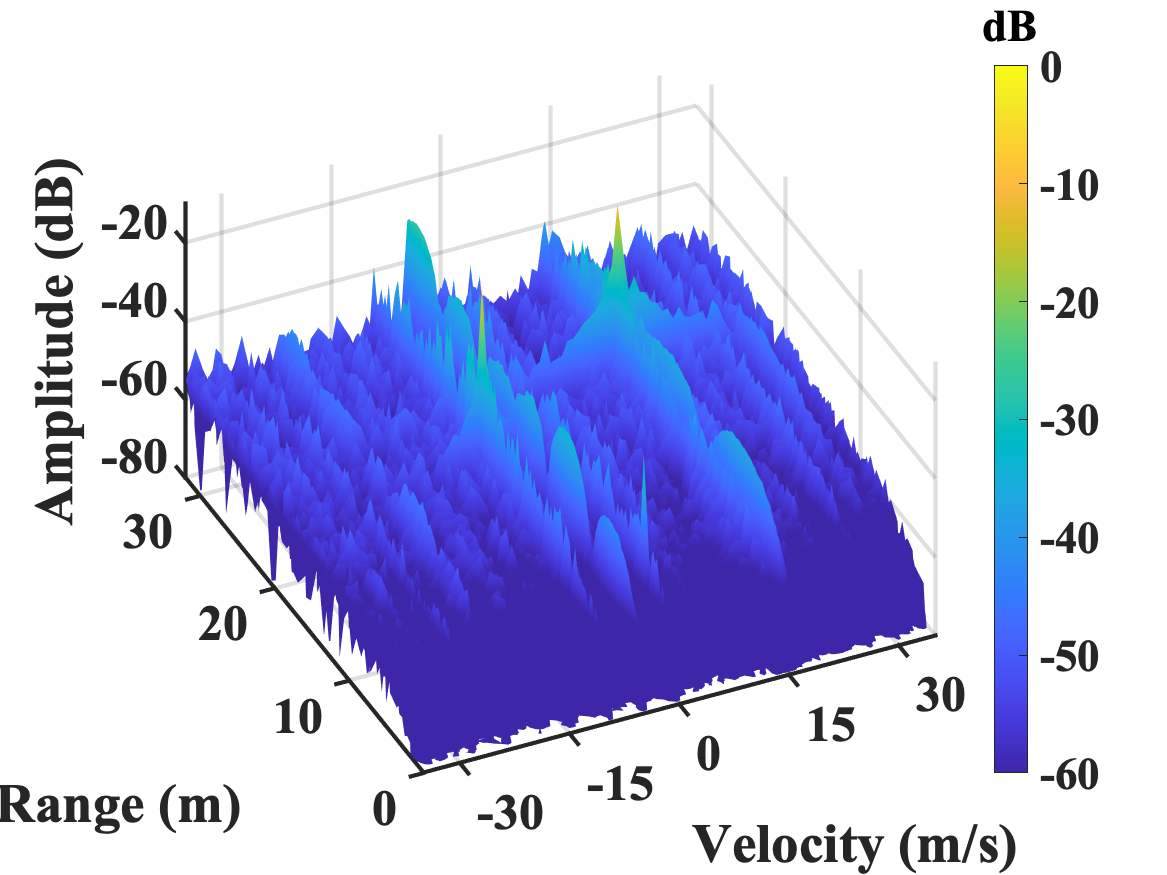}}
        \hspace{0mm}
	\subfloat[\label{rd:c}]{
		\includegraphics[scale=0.28]{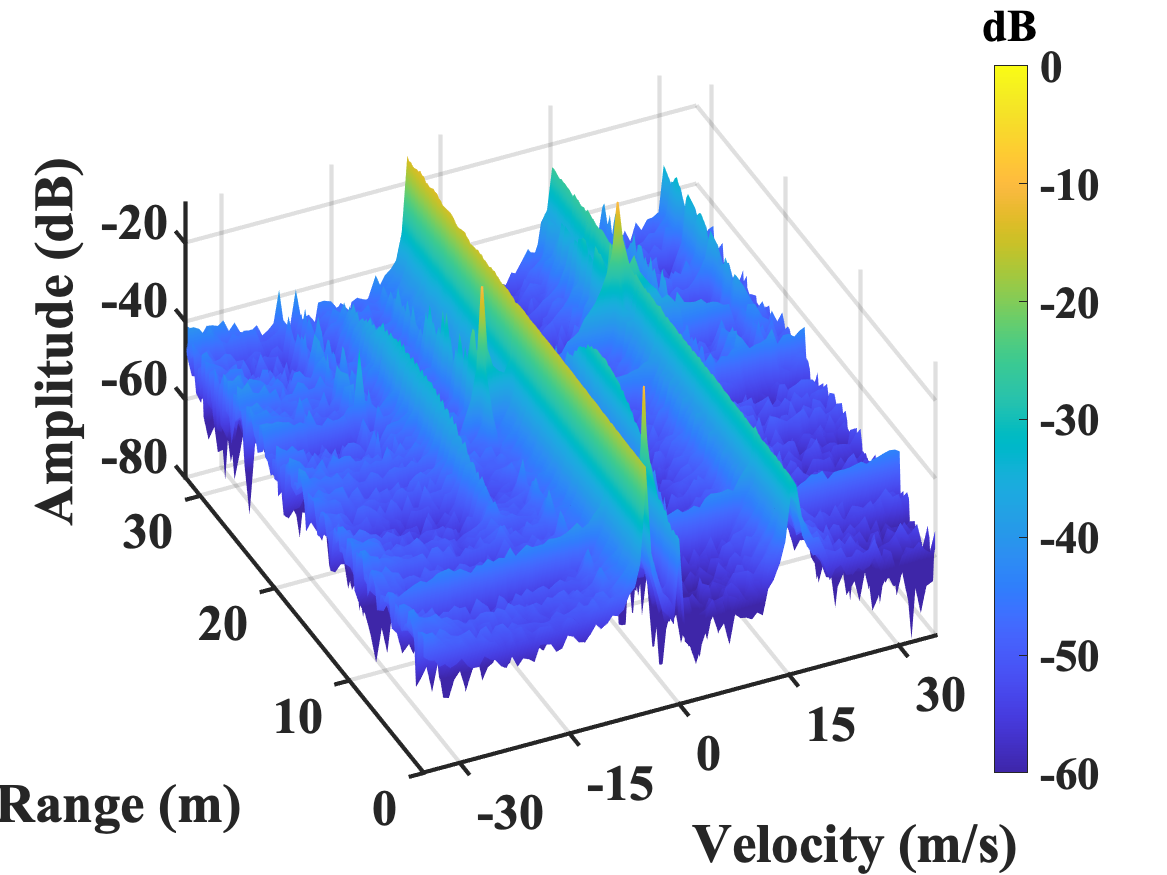}}
        \vspace{-2mm}
        \\
	\subfloat[\label{rd:d}]{
		\includegraphics[scale=0.28]{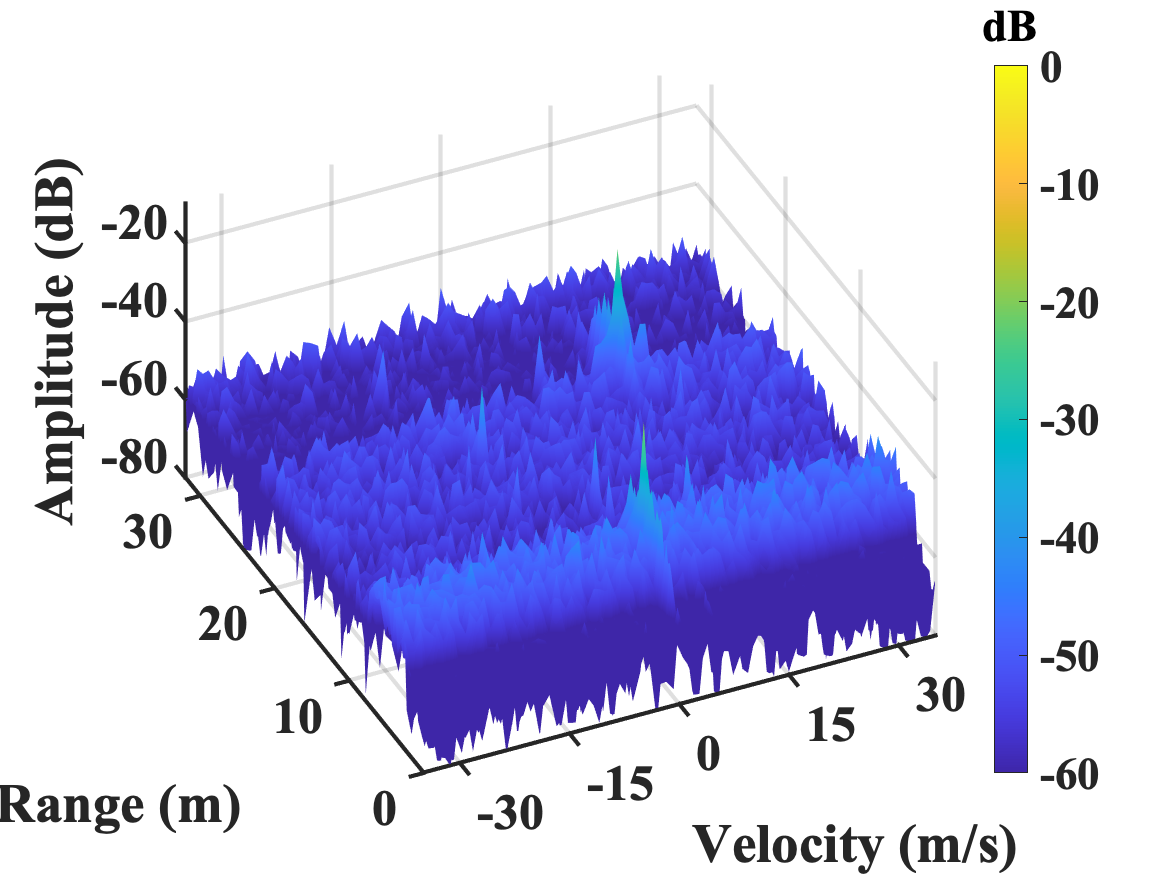} }
        \hspace{-1mm}
        \subfloat[\label{rd:e}]{
		\includegraphics[scale=0.28]{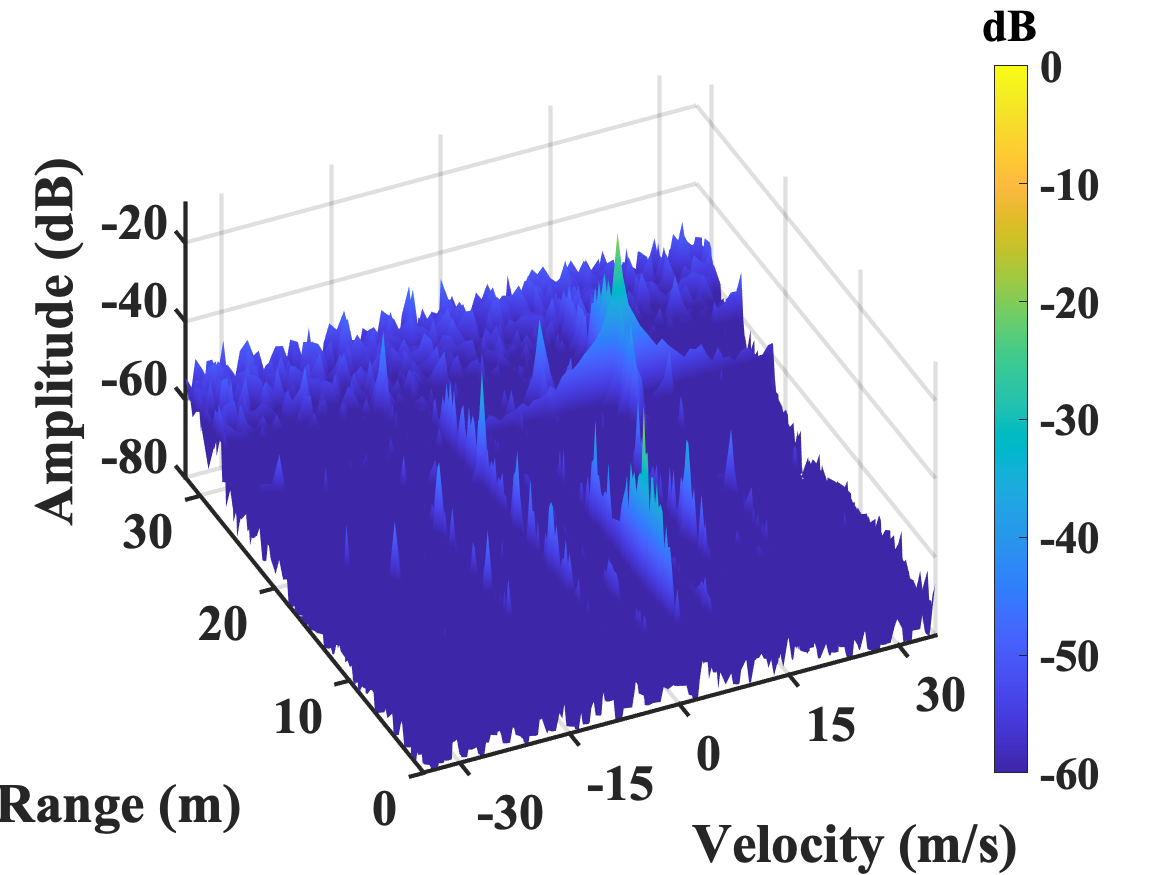} }
        \hspace{-1mm}
        \subfloat[\label{rd:f}]{
		\includegraphics[scale=0.28]{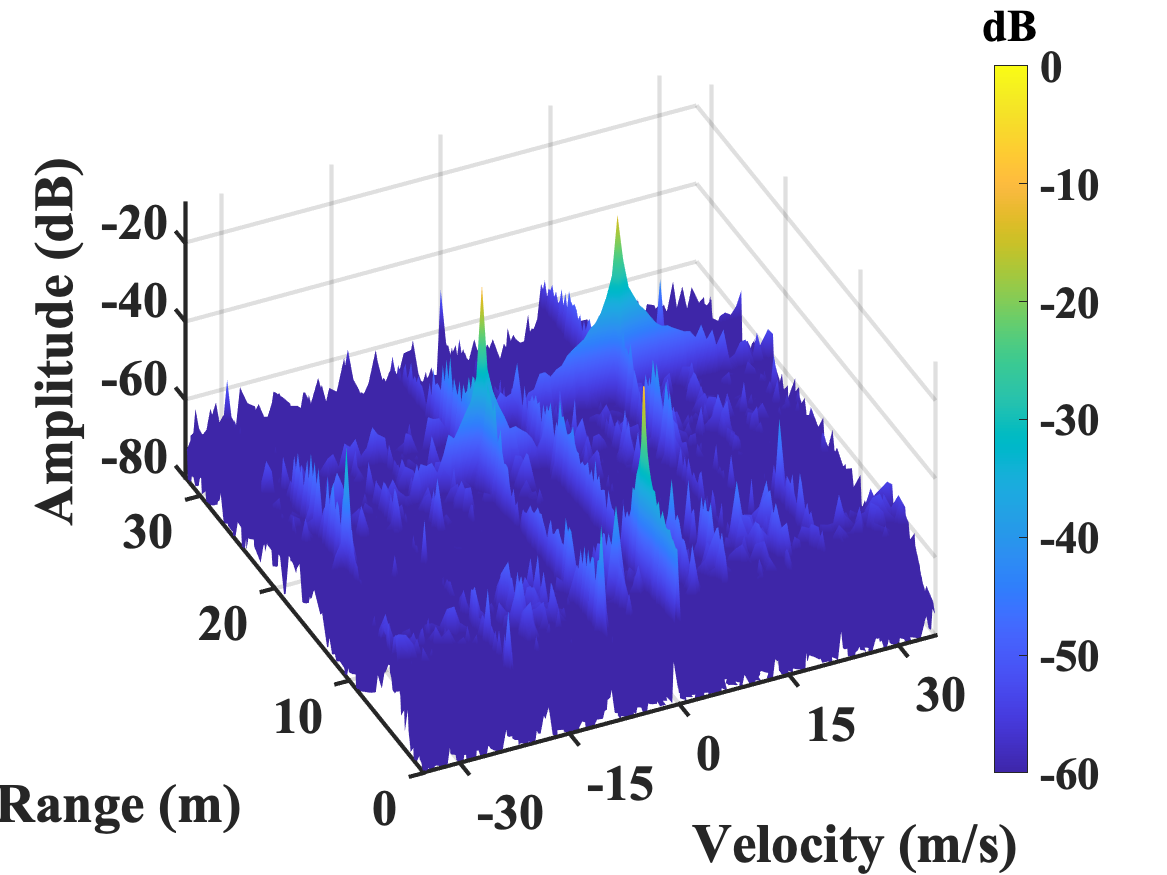} }
	\caption{Results produced by (a) signal with interference, (b) wavelet denoising, (c) ANC, (d) attention-based BiGRU, (d) complex-valued CNN, and (f) the proposed RIMformer on a set of 128 simulated chirps, presented as RD maps. }
	\label{rd}
\end{figure*}

\subsection{Performance Comparison}

\begin{figure}[tb]
\centerline{\includegraphics[scale=0.43]{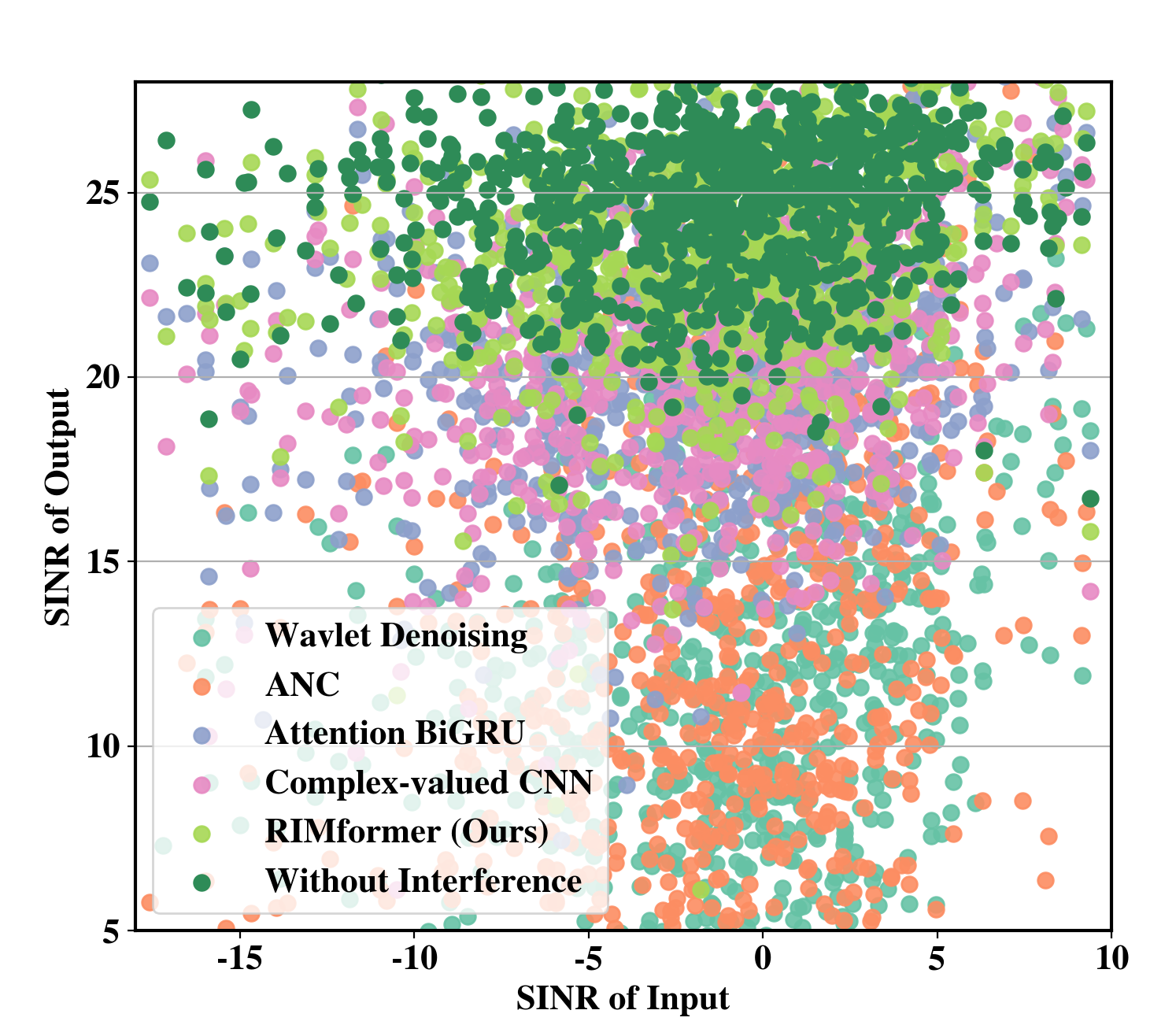}}
\caption{The target SINRs of different methods, including wavelet denoising, ANC, attention-based BiGRU, complex-valued CNN, the proposed RIMformer and the base signal without interference. Each point represents the target SINR of the interference mitigation result produced for one set of test data.}
\label{snr}
\end{figure}

To assess the effectiveness of the proposed method, a range of recently proposed techniques, including wavelet denoising~\cite{sp1}, ANC~\cite{sp2}, an attention-based bidirectional GRU (BiGRU)~\cite{dl1}, and a complex-valued CNN~\cite{dl5}, are chosen for conducting a comparative analysis. The latter three methods are trained for the same number of epochs based on simulated FMCW radar signals.

To illustrate the interference suppression effects of the tested methods, the results obtained for a simulated signal sample are depicted in both the time domain and the STFT domain in Fig.~\ref{stft}. Within the STFT domain, the spectrum of the target signal component is represented by a horizontal line along the time axis, while a slanted thick line signifies interference. The results show that the wavelet denoising method performs unsatisfactorily in the face of multiple severe interference factors. Due to the design principle of ANC, its T-F results retain the positive frequency part. The time-domain results show that a portion of the interference signal is suppressed. Both the attention-based BiGRU and complex-valued CNN can suppress interference. However, the amplitudes of the target signals are distorted. The proposed RIMformer recovers the target signals more accurately while eliminating the interference.

The results obtained for a set of 128 chirp signals are presented as range-Doppler (RD) maps in Fig.~\ref{rd}. The locations of spikes in the graph correspond to the distance and speed information of the targets in the simulated scenario. The values of the spikes indicate the intensity levels of the target echo. In this scenario, the speeds of the three targets are -7.63 m/s, 0 m/s and 18.31 m/s, while their distances are 3.5 m, 30.5 m and 16.7 m. A comparative analysis with different methods shows that the complex-valued CNN and the proposed RIMformer can achieve higher peaks and lower noise floors than those of the other approaches. The RIMformer is able to restore all three targets with an average amplitude of -18.36 dB, as shown in Fig.~\ref{rd} (f).

The results of the comparison conducted on the whole test set are represented as a scatter plot in Fig.~\ref{snr}. The horizontal and vertical coordinates represent the SINRs of the target in the input signal and the output results, respectively. Each point signifies the outcome produced for a single test sample after implementing interference mitigation, with the different methods distinguished by color. The greater proximity of a result on the plot to the clean signal indicates superior interference suppression. As illustrated in the figure, the proposed RIMformer obtains greater SINR values for the targets than do the four other methods.

\begin{table}[tb]
    \caption{Comparison Among the Results Produced by Different Methods}
    \begin{center}
    \begin{tabular}{cccc}
    \toprule
    \textbf{Method} & \textbf{Avg SINR$\uparrow$} & \textbf{Mid SINR$\uparrow$} & \textbf{Time $\downarrow$}\\
    \midrule
    Wavelet Denoising\cite{sp1} & 9.13 dB & 10.04 dB & \textbf{1.28 ms}\\
    ANC \cite{sp2} & 10.24 dB & 10.61 dB & \textbf{1.14 ms}\\
    Attention BiGRU\cite{dl1} & 20.68 dB & 21.32 dB & 211.16 ms\\
    Complex-Valued CNN\cite{dl5} & 21.49 dB & 22.67 dB & 6.83 ms\\
    \midrule
    \textbf{RIMformer (N=5)} & \textbf{23.12 dB} & \textbf{23.69 dB} & 14.46 ms\\
    \textbf{RIMformer (N=7)} & \textbf{24.25 dB} & \textbf{24.84 dB} & 19.37 ms\\
    \bottomrule
    \end{tabular}
    \label{tab1}
    \end{center}
\end{table}

Table~\ref{tab1} presents the numerical results of various methods and their corresponding processing times. Three variations of the RIMformer, each with different numbers of encoders and decoders, are compared. In line with our previous findings, a larger N parameter leads to superior results. With $N=7$, the RIMformer yields better results than those of the other comparison methods. In comparison with the network utilizing $N=5$, the incorporation of two additional encoders and decoders enhances the average SINR by 1.13 dB and introduces a time overhead of 4.91 ms. Compared with those of wavelet denoising, ANC, the attention-based BiGRU, and the complex-valued CNN, the average SINR of the proposed approach is 15.12 dB, 14.01 dB, 3.57 dB and 2.76 dB greater, respectively. The numerical SINR results validate the performance of the proposed RIMformer method in terms of mitigating interference in FMCW radar systems.

\section{Measurements}
\begin{figure*} [t!]
	\centering
	\subfloat[\label{fig:a}]{
		\includegraphics[scale=0.32]{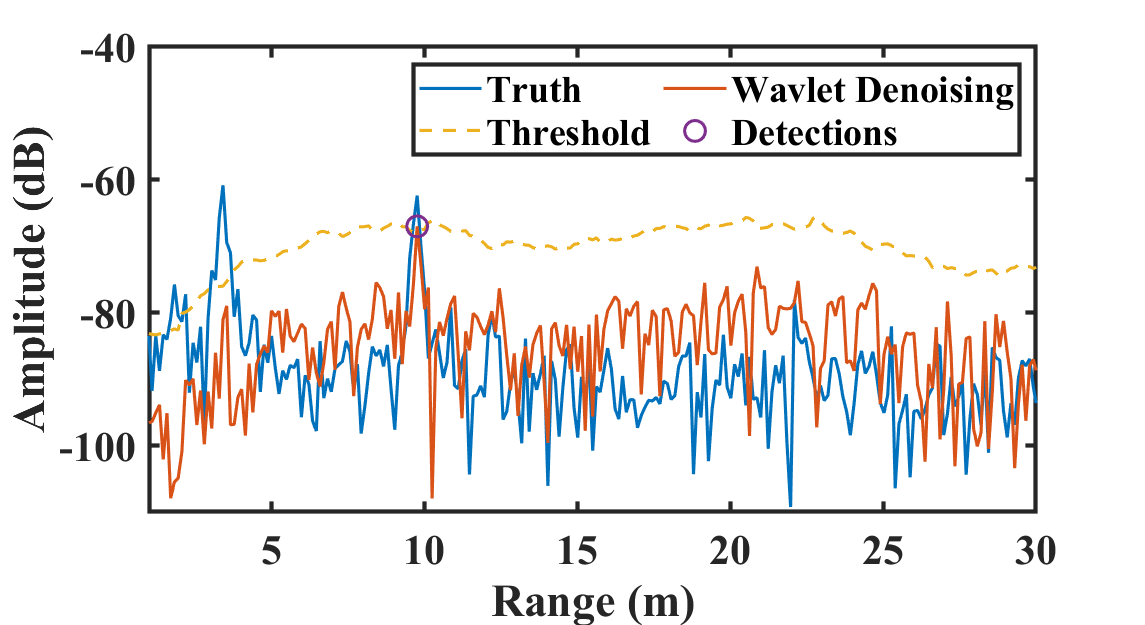}}
        \hspace{-7mm}
	\subfloat[\label{fig:b}]{
		\includegraphics[scale=0.32]{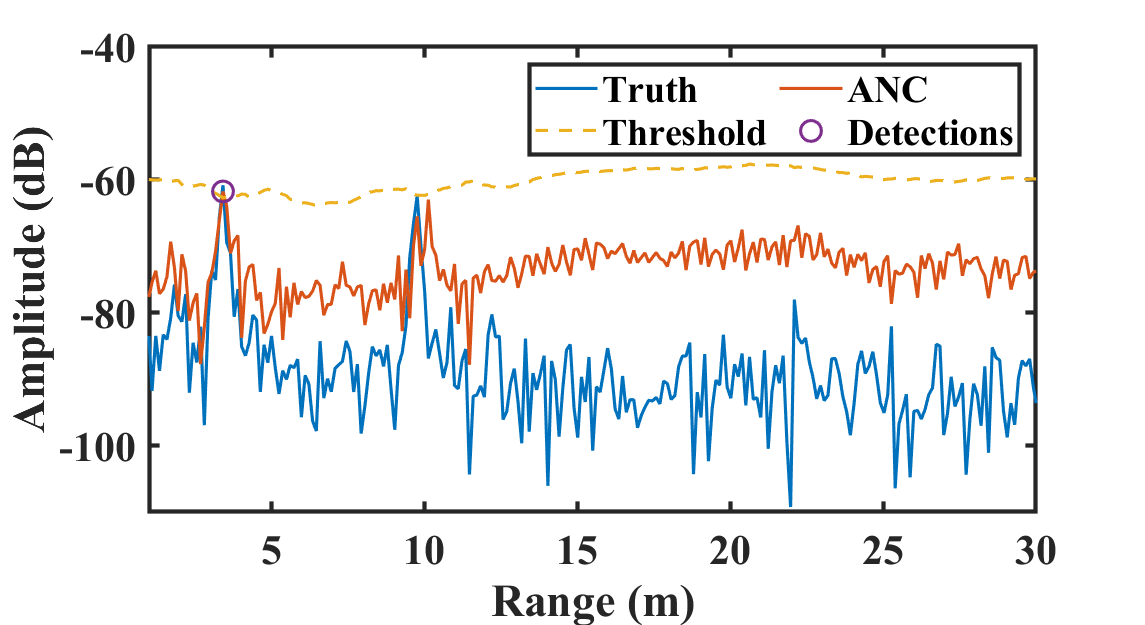}}
        \hspace{-7mm} 
	\subfloat[\label{fig:c}]{
		\includegraphics[scale=0.32]{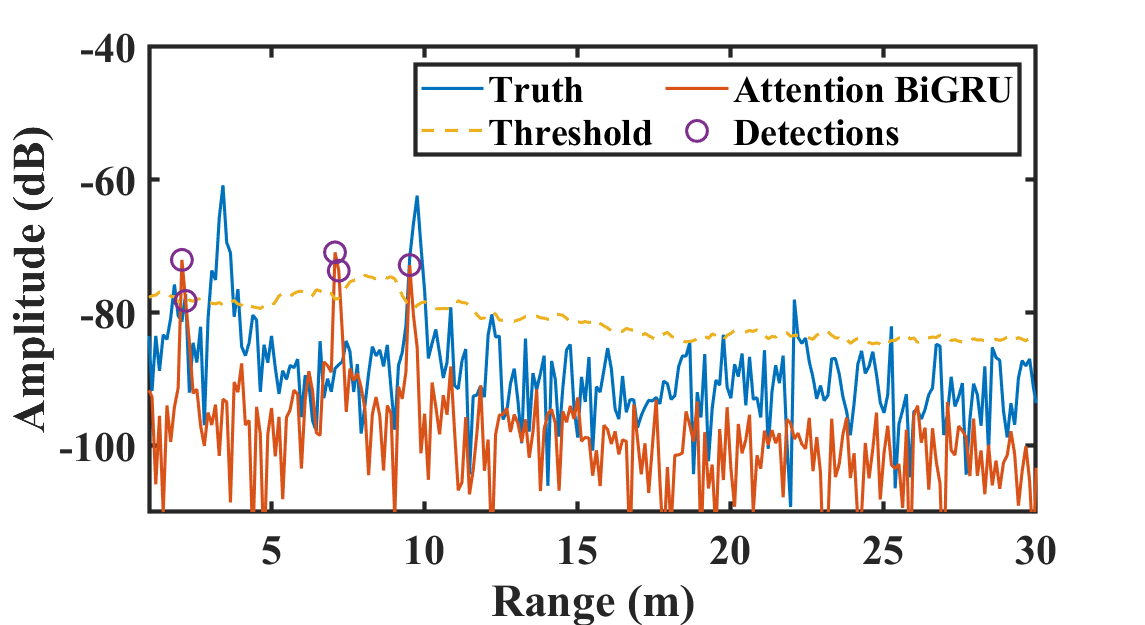}}
        \\
	\subfloat[\label{fig:d}]{
		\includegraphics[scale=0.32]{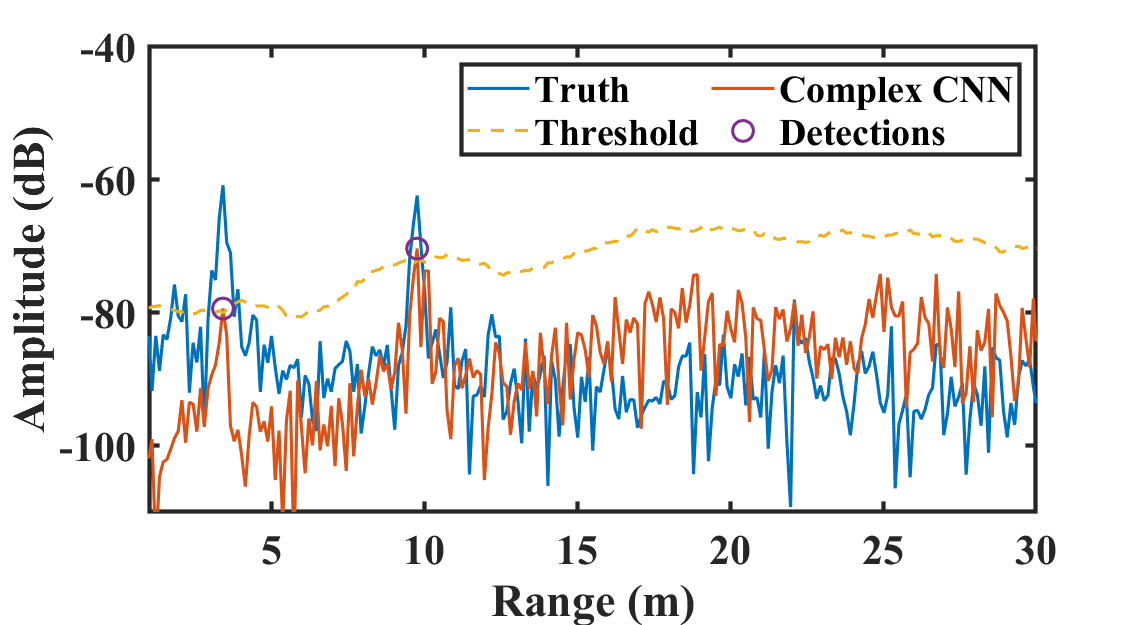} }
        \hspace{-7mm}
        \subfloat[\label{fig:e}]{
		\includegraphics[scale=0.32]{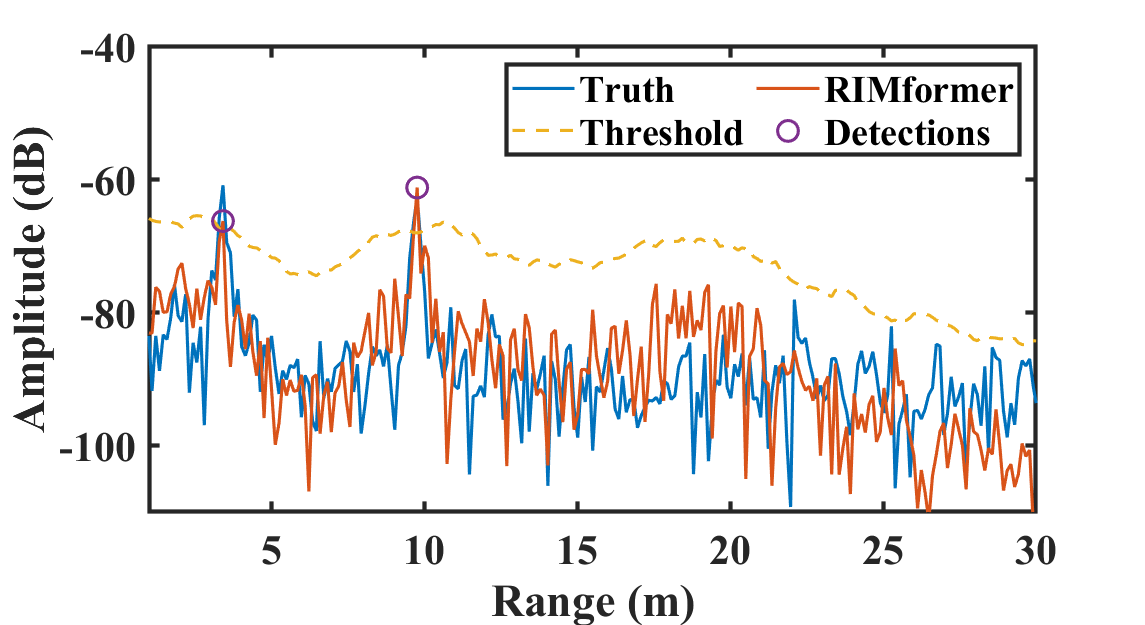} }

	\caption{CFAR detection results and spectrograms of the outputs obtained from (a) wavelet denoising, (b) ANC, (c) attention-based BiGRU, (d) complex-valued CNN, and (e) RIMformer on measured data.}
	\label{fft}
\end{figure*}

\begin{figure}[tb]
\centerline{\includegraphics[scale=0.6]{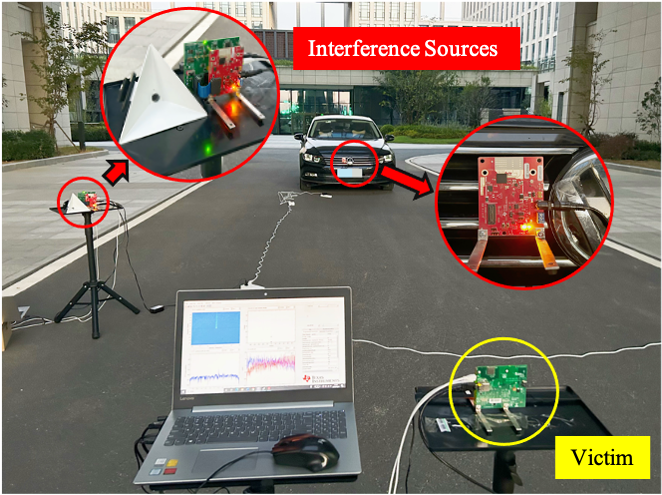}}
\caption{Scenario of the measured experiment.
The radar marked with a yellow circle in the bottom-right corner is the victim, while those marked with red circles are the interference sources.}
\label{photo}
\end{figure}

In this section, a measurement experiment is conducted to obtain data for validating the effectiveness of the proposed method in real-world scenarios. In the experiment, one FMCW mmWave radar serves as the victim, while the two other radars act as sources of interference. The experimental setup is illustrated in Fig.~\ref{photo}, where a nearby corner reflector and a distant car serve as the targets. The other two mmWave radars are positioned at the same locations as those of the targets, thus representing the radars equipped on the vehicle. The two targets are situated approximately 3 m and 10 m away from the receiving radar. The key parameters of both the victim and interference radars are detailed in Table~\ref{tab2}. The start frequencies of these radars are maintained at 77 GHz, with slope variations adjusted to 24 MHz/\textmu s, 10 MHz/\textmu s and 25 MHz/\textmu s. The interference radars with slopes similar to those of the victim radar generate blur with an extended time span, while the interference radar with substantial slope differences produces high-amplitude spikes.

\begin{table}[tb]
    \caption{Parameters of the Radars}
    \begin{center}
    \begin{tabular}{cccc}
    \toprule
    \textbf{Parameter} & \textbf{Victim} & \textbf{Interference 1} & \textbf{Interference 2}\\
    \midrule
    Start Frequency (GHz)& 77.0  & 77.0 & 77.0\\
    Frequency Slope (MHz/\textmu s)& 23.995 & 9.994 & 25.0\\
    Chirps per Frame & 128 & 128 & 32\\
    Frame Periodicity (ms)& 5.0 & 10.0& 100.0 \\
    ADC Samples & 1024 & -& - \\
    Sampling Rate (ksps) & 40000 & -& - \\
    \bottomrule
    \end{tabular}
    \label{tab2}
    \end{center}
\end{table}

The Fourier transform results of the outputs generated by the various methods are depicted in Fig.~\ref{fft}. It is evident that wavelet denoising exhibits a suboptimal recovery effect for the first target. Although the ANC method successfully recovers the nearest target, it introduces an offset for the second target, and the noise level at the far end of the spectrum remains elevated. The SINRs of the first two methods are relatively small. The BiGRU with attention yields a spurious spike between the two targets and induces an error when estimating the target distance. The complex-valued CNN introduces a significant bias when reducing the magnitude of the first target. In contrast, the RIMformer demonstrates better performance than that of the other methods in terms of restoring the targets.

To further evaluate the quality of the signals restored by the five interference mitigation methods, the target detection performance of the constructed ranging profiles is tested using the same constant false-alarm rate (CFAR) detector. Empirically, the threshold factor of the detector is set to 0.82. In addition to the resulting spectra yielded by the various methods, the results of the CFAR detector are also displayed in Fig.~\ref{fft}. All five methods detect one or more targets after employing RIM. Wavelet denoising detects the distant vehicle. The ANC method only detects the angle reflector at a close distance due to its high noise floor. The BiGRU method produces an incorrect target judgment. In contrast, the complex-valued CNN and RIMformer are able to correctly recognize all the targets and yield higher target detection probabilities. Moreover, the spectrogram produced by the RIMformer closely approximates the clean signal and reproduces the most accurate amplitude for the target. The difference between the predicted and real peaks is 1.21 dB. The experimental results obtained on measurement data and in a comparative analysis demonstrate the effectiveness of the RIMformer in real-world FMCW radar interference scenarios. The RIMformer preserves the integrity of the target signal while minimizing the amplitude distortion effect.

\section{Conclusion}

This paper introduces the RIMformer, a Transformer-based approach that is designed to mitigate interference from FMCW radar systems. The RIMformer seamlessly processes time-domain signals in an end-to-end manner, which eliminates the need for supplementary data processing or manual parameter tuning steps. By applying a dual multi-head self-attention mechanism, the RIMformer effectively captures the correlations among signal elements at different distances. The introduction of convolutional blocks allows the network to better extract localized features. A hybrid time-frequency loss function is proposed to learn the physical meaning of the target signal in the frequency domain. Through parameter optimization, the RIMformer strikes a balance between performance and computational efficiency. To validate the efficacy of the RIMformer, comprehensive comparisons are conducted with existing advanced methods. Both simulated and measured data are employed. The results demonstrate the superiority of the RIMformer in terms of suppressing interference and restoring the target signal.

\section{Acknowledgments}
The authors would like to thank Aoyong Dong, Lixiang He and Zhanhai He from the Hefei Innovation Research Institute of Beihang University for providing suggestions and help during the experiments.

\end{document}